\documentclass[pdflatex,sn-mathphys-num]{sn-jnl}

\usepackage{graphicx}%
\usepackage{multirow}%
\usepackage{amsmath,amssymb,amsfonts}%
\usepackage{amsthm}%
\usepackage{mathrsfs}%
\usepackage[title]{appendix}%
\usepackage{xcolor}%
\usepackage{textcomp}%
\usepackage{manyfoot}%
\usepackage{booktabs}%
\usepackage{algorithm}%
\usepackage{algorithmicx}%
\usepackage{algpseudocode}%
\usepackage{listings}%

\usepackage{mathtools}  

\usepackage{booktabs}
\usepackage{amssymb}
\usepackage{bbding}
\usepackage{pifont}
\usepackage{wasysym}
\usepackage{utfsym}
\usepackage{fontawesome}
\usepackage{mdframed}
\usepackage[table]{xcolor}
\usepackage{colortbl}

\usepackage{pgf}
\usepackage{tikz}
\usepackage[most]{tcolorbox}
\tcbset{enhanced,}

\usepackage{pgf}
\usepackage{pgfplots}
\pgfplotsset{compat=newest}

\usepackage{enumitem}
\usepackage{tcolorbox}
\tcbuselibrary{breakable} 

\usepackage{lipsum} 

\usepackage{titlesec} 
\usepackage{titletoc} 

\theoremstyle{thmstyleone}%
\theoremstyle{thmstyletwo}%

\theoremstyle{thmstylethree}%

\begin{document}

\title[Article Title]{Open World Knowledge Aided Single-Cell Foundation Model with Robust Cross-Modal Cell-Language Pre-training}


\author*[1,3,6]{\fnm{Haoran} \sur{Wang}}\email{wanghaoran@genomics.cn}

\author[1,2]{\fnm{Xuanyi} \sur{Zhang}}


\author[1]{\fnm{Shuangsang} \sur{Fang}}

\author[6]{\fnm{Longke} \sur{Ran}}

\author[1]{\fnm{Ziqing} \sur{Deng}}

\author[4,5]{\fnm{Yong} \sur{Zhang}}

\author[4,5]{\fnm{Yuxiang} \sur{Li}}

\author*[1]{\fnm{Shaoshuai} \sur{Li}}\email{lishaoshuai@genomics.cn}

\affil*[1]{\orgname{BGI Research}, \orgaddress{\state{Beijing}, \postcode{100083}, \country{China}}}

\affil*[2]{\orgname{School of Statistics}, \orgaddress{\orgname{Renmin University of China}, \state{Beijing}, \postcode{100872}, \country{China}}}

\affil*[3]{\orgname{BGI Research}, \orgaddress{\state{Chongqing}, \postcode{401329}, \country{China}}}

\affil*[4]{\orgname{BGI Research}, \orgaddress{\state{Wuhan}, \postcode{430074}, \country{China}}}

\affil*[5]{\orgname{BGI Research}, \orgaddress{\state{Shenzhen}, \postcode{518083}, \country{China}}}

\affil*[6]{\orgname{College of Artificial Intelligence Medicine}, \orgaddress{\orgname{Chongqing Medical University}, \postcode{400030}, \state{Chongqing}, \country{China}}}




\abstract{
Recent advancements in single-cell multi-omics, particularly RNA-seq, have provided profound insights into cellular heterogeneity and gene regulation. While pre-trained language model (PLM) paradigm based single-cell foundation models have shown promise, they remain constrained by insufficient integration of in-depth individual profiles and neglecting the influence of noise within multi-modal data. To address both issues, we propose an Open-world Language Knowledge-Aided Robust Single-Cell Foundation Model (OKR-CELL). It is built based on a cross-modal Cell-Language pre-training framework, which comprises two key innovations: (1) leveraging Large Language Models (LLMs) based workflow with retrieval-augmented generation (RAG) enriches cell textual descriptions using open-world knowledge; (2) devising a Cross-modal Robust Alignment (CRA) objective that incorporates sample reliability assessment, curriculum learning, and coupled momentum contrastive learning to strengthen the model's resistance to noisy data. After pretraining on 32M cell-text pairs, OKR-CELL obtains cutting-edge results across 6 evaluation tasks. Beyond standard benchmarks such as cell clustering, cell-type annotation, batch-effect correction, and few-shot annotation, the model also demonstrates superior performance in broader multi-modal applications, including zero-shot cell-type annotation and bidirectional cell-text retrieval.
}




\maketitle

\section{Introduction}\label{sec1}


Recent progress in single-cell multi-omics technologies \cite{chappell2018single,hu2024benchmarking,badia2023gene,kartha2022functional,lin2022clustering}, particularly high-resolution RNA-seq, have revolutionized our understanding of cellular heterogeneity and gene regulatory mechanisms, laying the foundation for critical applications such as drug response prediction \cite{2024The} and disease mechanism dissection. Parallel to this, the success of Pre-trained Language Models (PLMs) \cite{devlin2019bert,radford2018improving,brown2020language} has spurred the development of single-cell foundation models \cite{lopez2018deep,yang2022scbert,theodoris2023transfer,hao2024large,cui2024scgpt}, which aim to distill generalizable biological insights from large-scale transcriptomic data. However, most of them rely solely on uni-modal information, failing to capture the complex, multi-faceted biological characteristics inherent to cells.


Language serves as a carrier of human biological knowledge, enabling foundation models to decode the "life language" contained in cells. While early attempts like BioTranslator \cite{xu2023multilingual} leveraged text-only pretraining, recent cross-modal Cell-Text Pretraining (CTP) methods \cite{zhao2024langcell, bian2024scmulan} have advanced by adopting cross-modal alignment frameworks \cite{radford2021learning,li2022blip,wang2020consensus,li2020oscar,wang2022coder}. For instance, LangCell \cite{zhao2024langcell} integrated cell-text cross-attention and contrastive learning to transfer textual knowledge, and scMULAN \cite{bian2024scmulan} constructed "c-sentences" by fusing transcriptomics with meta-textual attributes. 

Despite these strides, current CTP approaches suffer from two critical and unresolved limitations: (1) \textbf{Impoverished and constrained textual corpora}: The original textual data primarily consists of isolated attribute descriptors (e.g. cell type labels) that lack contextual dependencies among them. For example, macrophages function as microglia in the brain and Kupffer cells in the liver nuances unconveyed by mere labels, resulting in superficial cellular state understanding compared to gene-level transcriptomic profiles. Furthermore, these specialized corpora are confined to the knowledge scope of their creators, failing to incorporate the vast, evolving open-world biological knowledge essential for comprehensive cellular interpretation. (2) \textbf{Vulnerability to multi-modal noise}: Real-world cell-text data is inherently noisy, yet traditional CTP methods (e.g., CLIP-based frameworks) ignore this problem. In the cellular modality, technical variability across sequencing platforms/protocols introduces gene dropout and batch effects \cite{klein2015droplet,liu2024sequencing}; in the textual modality, phenotypic descriptions are often subjective, incomplete, or inaccurate due to varying biological expertise \cite{ma2025tissue}. Such noise-induced misalignments severely undermine the fidelity of cross-modal learning, restricting model's robustness and generalization.

To address these challenges, we propose an \textbf{O}pen-World Language \textbf{K}nowledge-Aided \textbf{R}obust Single-\textbf{Cell} Foundation Model (\textbf{OKR-CELL)}, which integrates open-world biological knowledge and robust cross-modal alignment into a unified framework. Our core innovations are two-fold:

First, at the data level, we augment textual corpora using Large Language Models (LLMs) enriched with retrieval-augmented generation (RAG). Leveraging LLMs’ open-world knowledge, we generate comprehensive, context-rich textual descriptions for each cell—going beyond static attributes. Meanwhile, to mitigate LLM hallucinations, we retrieve  specialized bioinformatics literature databases (\textit{e.g.}  \cite{canese2013pubmed}) based on original metadata, ensuring the generated text is biologically reliable through RAG.  

Second, at the method level, we design a novel Cross-modal Robust Alignment (CRA) objective. Unlike conventional contrastive learning that blindly maximizes positive/negative sample discrimination, CRA explicitly accounts for noise from four aspects: (i) For positive sample pairs, we introduce the concept of positive sample pair reliability to emphasize highly reliable positive pairs and suppress low-reliability ones; (ii) For negative pairs, we enhance discriminability across samples of varying reliability by indirectly optimizing their matching probabilities, which prioritizes low-probability (high-confidence) negatives while suppressing the influence of high-probability (potentially false negative) ones. (iii) To alleviate early-stage reliability assessment limitations on negatives, a curriculum learning based strategy is presented to progressively increase the emphasis on negative sample pairs. (iv) We introduce negative sample queues to expand candidate negatives to promote cross-modal contrastive learning.

OKR-CELL adopts a multi-task learning framework, which leverages scGPT as backbone to process RNA-seq data and Clinical-LongFormer as the textual encoder. On one hand, we learn intra-modal cellular information using scGPT’s masked gene modeling objective; On the other hand, we align cell and textual representations in a shared embedding space to convey complementary information across modalities. To train this model, we construct a SCxGEN-32M dataset, consisting of 32 million cell-text pairs crawled from CELLXGENE \cite{czi2025cz} platform, each with 9 key metadata items (cell type, tissue origin, etc.). Extensive experiments demonstrate that OKR-CELL achieves state-of-the-art (SoTA) performance across 6 tasks: In traditional tasks, we achieve superior performance in cell clustering, cell type annotation, and batch effect correction. Furthermore, we extend our evaluation to multi-modal cell understanding tasks. Our model also exhibits superior performance in zero-shot cell type annotation and bidirectional cell–text cross-modal retrieval, underscoring its robust generalization to previously unseen cell classes.

Our main contributions can be summarized as follows:
\begin{itemize}
	\item We propose OKR-CELL, the first single-cell foundation model that integrates LLM-derived open-world knowledge to enable deep, individualized cellular understanding.
	
	\item We devise a novel Cross-modal Robust Alignment (CRA) objective, which leverages complementary learning to improve the robustness of model to the noisy training data. Moreover, a Coupled Momentum Constrastive Learning and Curriculum Learning based strategy are incorporated into CRA loss to enhance cross-modal alignment. 

    \item The extensive experiments on multiple benchmark datasets validate the superiority of our model in both traditional biological tasks and multi-modal understanding, advancing its generalization to real-world noisy scenarios.
\end{itemize}

\section{Results}\label{sec2}

\subsection{Method Overview}

\subsubsection{Model Architecture}
\begin{figure*}[t]
    \begin{center}
\includegraphics[width=1\linewidth]{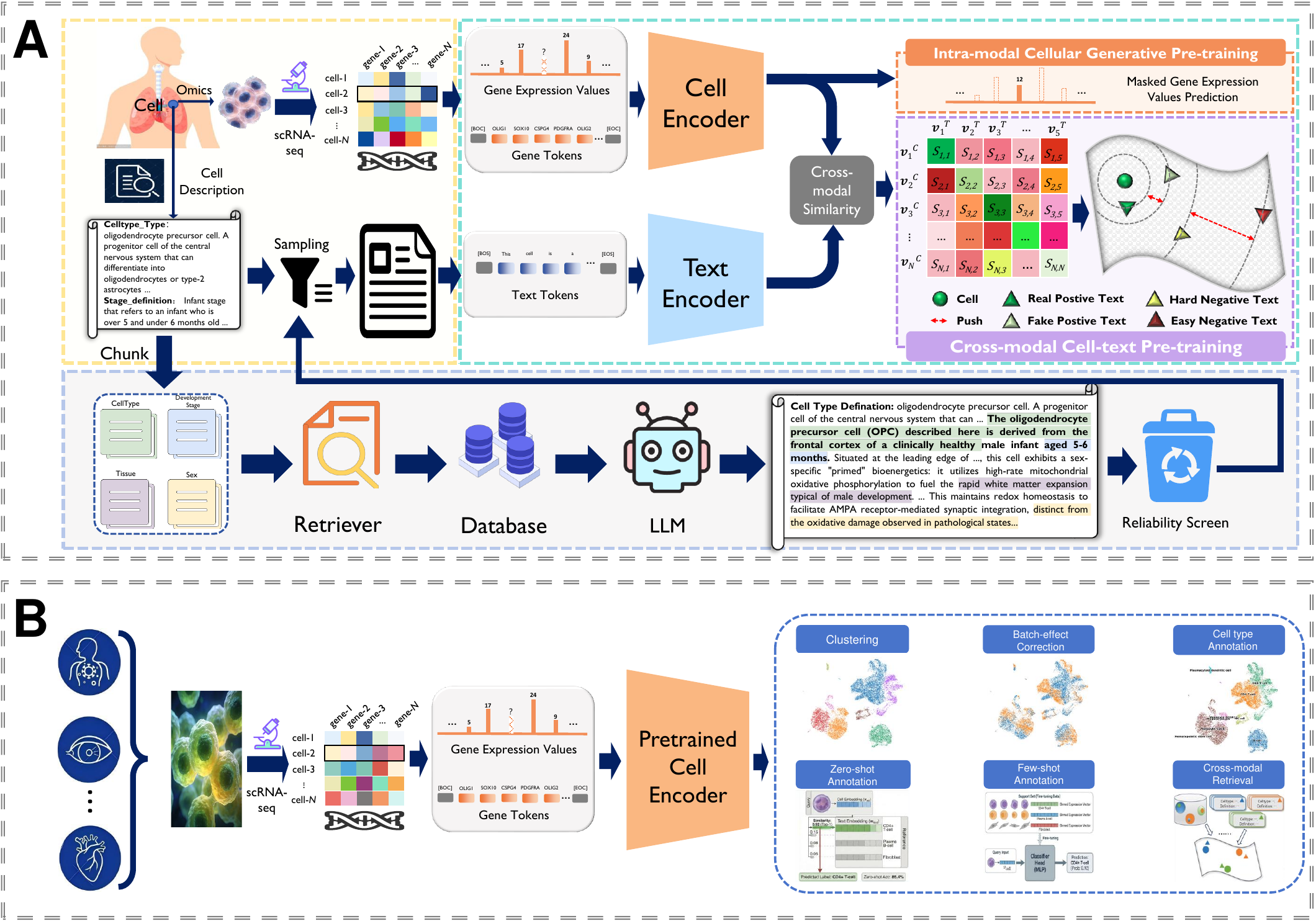}
    \end{center}
    \caption{(A) The schematic overview of the OKR-CELL method. (B) The illustration of several downstream tasks implemented via OKR-CELL, including cell clustering, batch affect correlation, cell-type annotation and cross-modal retrieval.}
    \label{fig:0}
\end{figure*}

In this section, we present a comprehensive description about the workflow of our OKR-CELL method, whose framework is illustrated in Figure \ref{fig:0}, OKR-CELL is composed of three trainable components: a cell encoder, a text encoder and a cross-modal projector connecting cellular and textual modalities. 1) For cell encoder, we employ utilize a transformer-based single-cell foundation model, scGPT \cite{cui2024scgpt} to handle the gene tokens and expression values that are specific to scRNA-seq data. 2) For the textual branch, a longformer architecture based clinical corpora enriched language model named Clinical-Longformer \cite{li2022clinical} is utilized, whose maximum input sequence token length is 4,096. It is pre-trained on approximately 2 million clinical notes extracted from the MIMIC-III dataset. 3) Cross-modal projectors. We employ a cell-to-text connector to map cellular representation into a joint space consistent with textual representation. 

\subsubsection{Pre-training Objective and Pipeline}

Our OKR-Cell model is constructed by projecting scRNA-seq data and text into a common latent space. The pretraining objective of our model includes two parts: 1) One part is Intra-Modal Generative Pre-training (IMGP) objective on single-cell transcriptomics domain, including Gene Expression Prediction (GEP), Gene Expression Prediction for Cell Modelling (GEPC) and Cell Type Classification (CLS) (see Section \ref{ICGP}). 2) The other part is Cell-Text Cross-modal Robust Alignment (CRA) objective (see Section \ref{CRA}), which leverages the high-level knowledge contained in natural language as guidance to assist cell representation learning. 

Our model is achieved by two-stage pre-training. A two-stage pre-training paradigm is adopted for our model, trained based on the SCxGEN-32M dataset (see details in Section \ref{Supp_DatasetStat}) along with the textual corpus enriched by LLM (see details in Section \ref{sec_Data}). On the first stage, we aim to establish basic alignment between cellular data and textual representations. Thus we freeze the parameters of textual encoder and optimize the parameters of the cell encoder and the cell-to-text projector. Based on the alignment between cellular and textual modality after stage-1, on the second stage, we unfreeze the textual encoder to further facilitate mutual propagation of information between the two modalities. Both IMGP and CMA objectives are employed on the two stages.

\subsection{Comprehensive Evaluation of The Cell Representation Capacity of scFMs}

For a single-cell foundation model, cell type clustering analysis is a common criteria to validate its ability of cell information understanding. We first evaluate the clustering performance under unsupervised setting on four datasets: blood \cite{de2025comprehensive}, kidney \cite{jorstad2023comparative}, hapancreas \cite{chen2023transformer}, and hpbmc \cite{tran2020benchmark}. For performance comparison, we employ four prevailing single-cell foundation models for comparison: Geneformer \cite{theodoris2023transfer}, scBERT \cite{yang2022scbert}, scFoudantion \cite{hao2024large}, scGPT \cite{cui2024scgpt}, which are all trained on single-modal scRNA-seq data. Besides, we compare our OKR-CELL with two cross-modal cell-language pre-training based methods: LangCell \cite{zhao2024langcell} and scCLIP-GPT (see details in \ref{comMethod}). 

\begin{figure*}[t]
    \begin{center}
\includegraphics[width=1\linewidth]{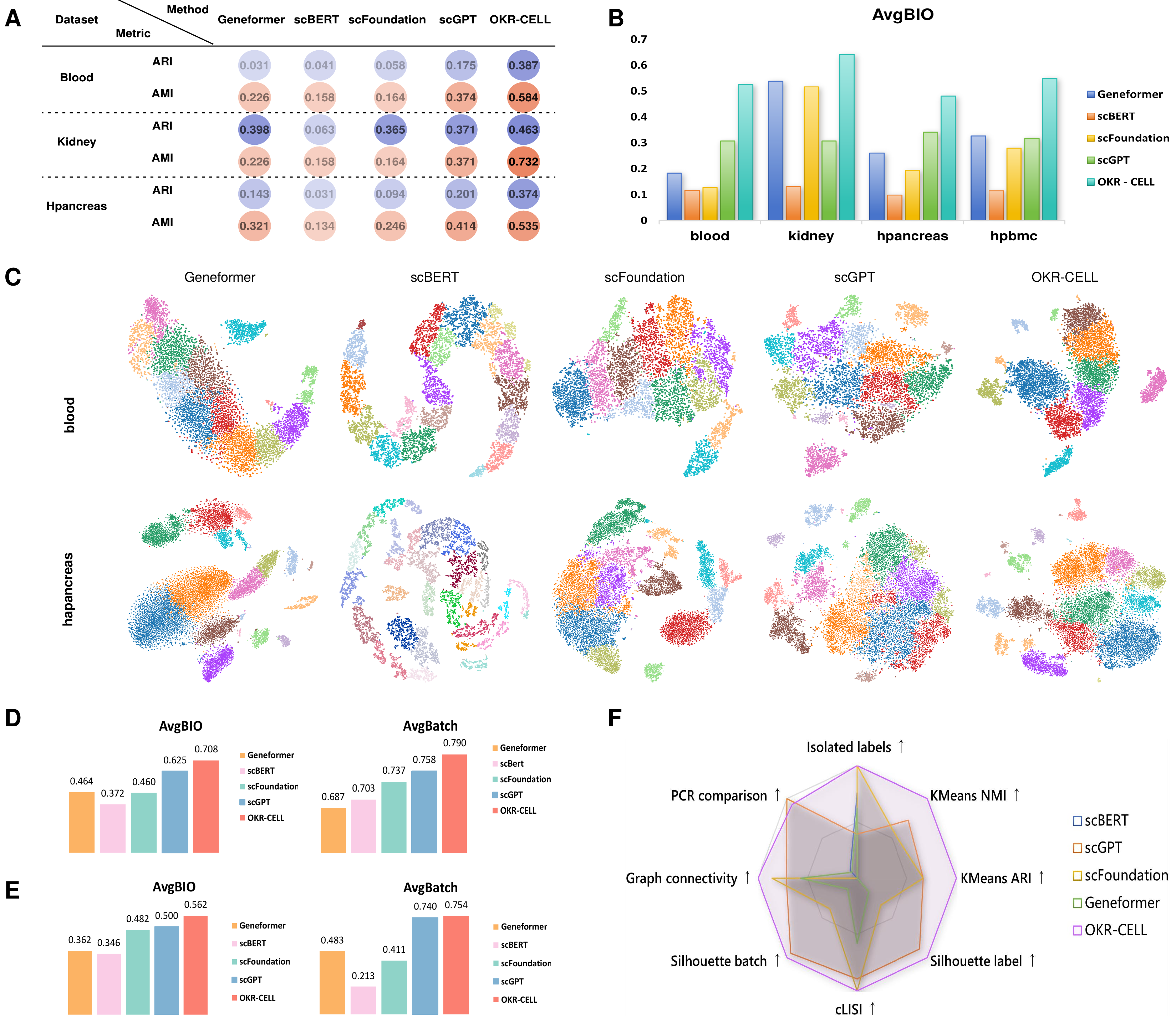}
    \end{center}
    \caption{(A) Summary table of ARI and AMI scores for cell clustering across three datasets (blood, kidney, and hPancreas). (B) Bar plot of AvgBIO scores across four datasets (blood, kidney, hPancreas and hpbmc). (C) T-SNE plots of cell embeddings generated by different methods on blood and hPancreas datasets. (D) Bar plot of AvgBIO and AvgBatch scores of batch effect correction by different methods on hPancreas dataset. (E) Bar plot of AvgBIO and AvgBatch scores of batch effect correction by different methods on hpbmc reference dataset. (F) Radar chart of 8 cretria evaluating the effectiveness of batch effect correction by different methods on hPancreas reference dataset. }
    \label{fig:1}
\end{figure*}

OKR-CELL demonstrates superior clustering capabilities across three benchmark datasets (Blood, Kidney, and Hpancreas), outperforming competing models (Geneformer, scBERT, scFoundation, and scGPT) in both quantitative metrics and biological interpretability. Seen from Figure \ref{fig:1}A, OKR-CELL achieves the highest ARI (Adjusted Rand Index) and AMI (Adjusted Mutual Information) scores consistently across all datasets: on the Blood dataset, it reaches an ARI of 0.387 and AMI of 0.584, which are 1.2 and 2.2 times higher than the second-best performer scGPT (ARI=0.175, AMI=0.374); on the Kidney dataset, its AMI of 0.732 stands out as the most prominent, surpassing scGPT’s 0.371 by nearly double; even on the Hpancreas dataset, where overall clustering performance is constrained, OKR-CELL maintains the lead with an ARI of 0.374 and AMI of 0.535. To achieve more comprehensive evaluation on clustering results, we calculate AvgBIO scores (see Section \ref{sec_AvgBIO}) of comparison methods on all four datasets, which measures the retention level of genuine biological signals. Figure \ref{fig:1}B indicates that OKR-CELL effectively captures tissue-specific biological structures. For instance, distinguishing rare cell subtypes in blood and resolving heterogeneous cell populations in kidney tissue—that other models fail to prioritize, as reflected by their substantially lower metric scores. Visualization results of data distribution from Figure \ref{fig:1}C further confirm that OKR-CELL’s clustering outputs align more closely with known biological annotations, validating its ability to translate quantitative gains into meaningful biological insights.

As a major confounder for cell type clustering is the inclusion of multiple datasets from different sequencing batches or technologies, batch effect is a more challenging task to testify the representation ability of single-cell foundation models. To assess the model's ability on batch effect correction, we conduct corresponding experiments on hapancreas and hpbmc datasets. As for metrics, besides AvgBIO score, AvgBatch score (see Section \ref{sec_AvgBatch}) is also computed to reflect the elimination degree of non-biological noise derived from batch variations. From Figure \ref{fig:1}D, on hapancreas dataset, we can see OKR-CELL achieves the highest AvgBIO score (0.708), a key metric for balancing batch removal and biological signal retention, outperforming scGPT (0.687); it also leads in AvgBatch score (0.790), indicating stronger integration of cross-batch cells without obscuring true biological differences. Figure \ref{fig:1}E shows OKR-CELL also exhibits best performance on both two metics on the hpbmc dataset. Additionally, in the Figure \ref{fig:1}F, we demonstrate eight specific indicators (see Section \ref{sec_batchEffect}) to more comprehensively and meticulously evaluate the ability of various methods to eliminate batch effects from multiple aspects. It can be seen that our OKR-CELL performs best on all indicators with the sole exception of PCR index. Overall, our OKR-CELL excels coping with the noise induced by batch effects in sequencing technique while preserving biological heterogeneity, as evidenced by multiple complementary metrics across datasets.

\subsection{Cell Type Annotation}
Besides clustering in single-cell analysis, precise cell type annotation serves as the cornerstone for large-scale single-cell RNA sequencing (scRNA-seq) analyses. It deciphers the heterogeneity inherent in sequenced tissues and establishes a fundamental framework for subsequent explorations into cellular and gene functions, thereby facilitating the acquisition of biological and pathological insights. For traditional cell type annotation task, we also employ four aforementioned methods for comparison: Geneformer, scBERT, scFoudantion, scGPT.  

Moreover, in real-world applications, acquiring sufficient high-quality labeled data for each target cell type during the fine-tuning process remains a significant hurdle. This poses a critical challenge to the practical deployment of existing single-cell models. To tackle this problem, we present to introduce few-shot learning and zero-shot learning into benchmark against baselines and evaluate the data efficiency of models. These settings allows us to evaluate the cell type annotation capability of single-cell foundation models under scenarios where the number of classes in the test samples is extremely few or even unseen. Under these settings, we select LangCell and scCLIP-GPT for comparison. Following previous studies, the utilized evaluation metrics are accuracy and macro F1-score (\textbf{abbreviated as F1-Score} in the rest manuscript) for all cell type annotation tasks.

\subsubsection{Traditional Classification}
\begin{figure*}[h!]
    \begin{center}
\includegraphics[width=1\linewidth]{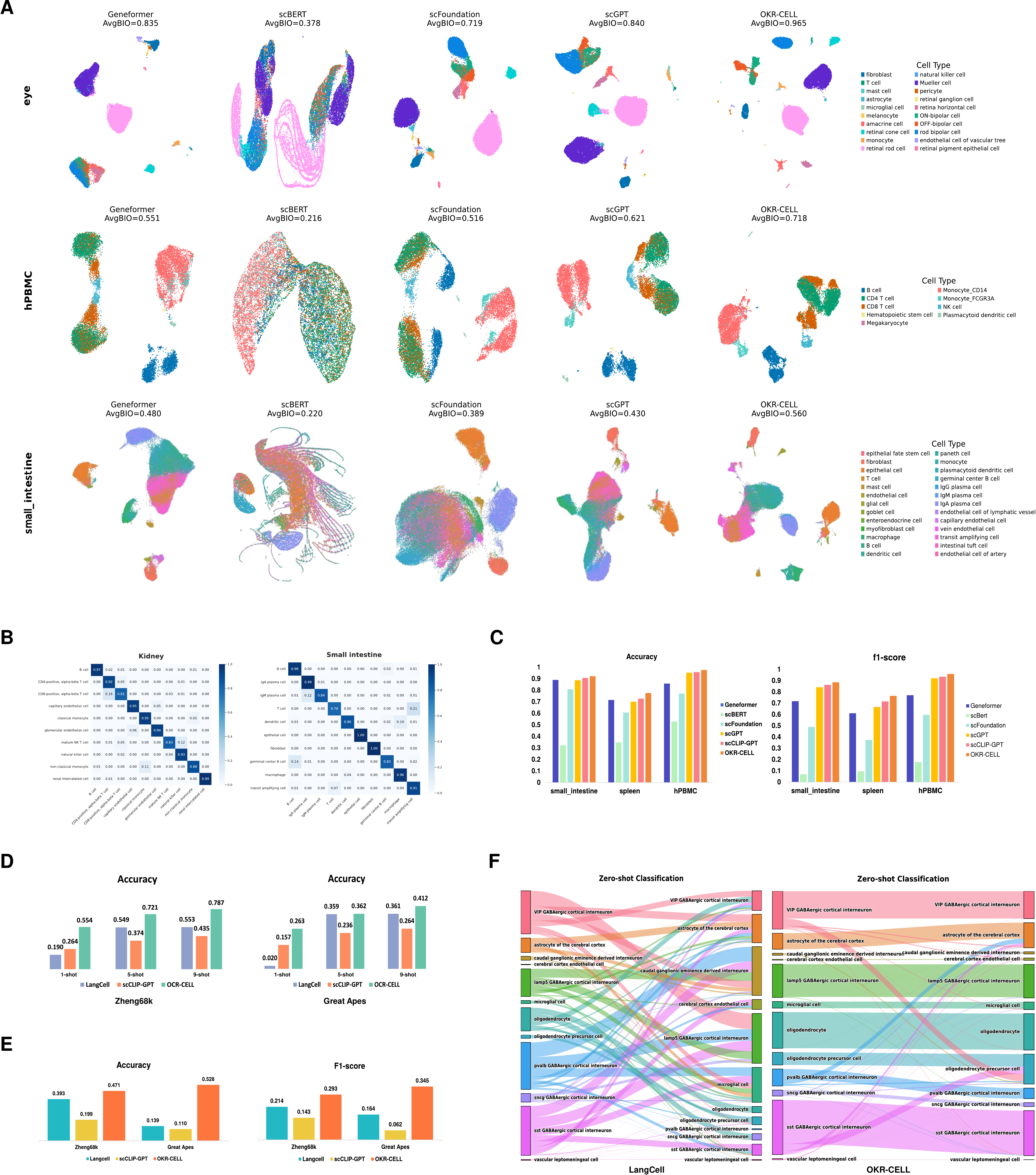}
    \end{center}
    \caption{(A) T-SNE visualization plots of cell embeddings generated by different methods on eye, hpbmc and small-intestine reference datasets colored by ground-truth cell types. (B) Normalized confusion matrix by cell types on kidney and small-intestine reference datasets by OKR-CELL method. (C) Bar plot of accuracy and f1-score for cell type annotation by different methods on small-intestine, spleen and hpbmc reference datasets. (D) Bar plot of accuracy and f1-score for few-shot cell type annotation different methods on zheng68k and Great-Apes reference datasets. (E) Bar plot of accuracy and f1-score for zero-shot cell type annotation different methods on eye, prostate-gland and Great-Apes reference datasets. (F) Sankey diagram  visualizes the correspondence between predicted labels and reference annotations for zero-shot cell type annotation by LangCell and OKR-CELL methods on prostate-gland reference dataset.}
    \label{fig:2}
\end{figure*}

The classical cell type annotation task refers to finetuning the pre-trained foundation model on tissue-specific dataset and validate its in-domain classification performance. OKR-CELL demonstrates exceptional performance in traditional cell type annotation across multiple tissue-specific datasets, outperforming competing models in both accuracy and macro f1-score metrics. We first conduct the clustering analysis on cell embeddings marked with true cell type label. From Figure \ref{fig:2}A, we can see OKR-CELL achieves the highest AvgBIO score over all three datasets. Concretely, on the eye \cite{cowan2020cell} dataset, OKR-CELL is able to effectively distinguish among ON-bipolar cells, OFF-bipolar cells, and rod bipolar cells while well preserving the semantic relationships between the three cell types. On the hpbmc dataset, the confusion degree in the distribution between CD8 T cells and CD4 T cells obtained using the OKR-CELL model is the lowest among all methods, which possess highly similar biological functions. 

To evaluate the discriminative quality of cell representations, we performed Logistic Regression analysis on the Kidney and Small Intestine \cite{zheng2021concerted} datasets. As shown in Figure \ref{fig:2}B, the embeddings exhibited high linear separability, with diagonal accuracy exceeding 0.95 for majority classes. Importantly, misclassifications in the confusion matrices are biologically meaningful rather than random. Specifically, in the Kidney dataset, overlap is primarily observed between functionally related types, such as CD8+ and CD4+ T cells (16\%), reflecting their shared transcriptomic signatures. Similarly, in the Small Intestine dataset, while structural cells like Fibroblasts achieved 100\% accuracy, minor confusion was confined to developmental lineages (e.g., 12\% of IgM plasma cells misclassified as IgA plasma cells). These results confirm that OKR-CELL captures distinct cellular identities while preserving hierarchical biological semantics.

To rigorously evaluate classification performance, we benchmarked OKR-CELL against five representative foundation models, including the enhanced baseline scCLIP-GPT, across three organ-specific datasets. As shown in Figure \ref{fig:2}C, OKR-CELL consistently surpasses all competing methods in both accuracy and F1-score. Notably, even in comparison with the strong scCLIP-GPT baseline, OKR-CELL demonstrates a clear and consistent advantage, which is especially pronounced in the challenging spleen \cite{xu2023automatic} dataset, where it attains a substantially higher F1-score. Moreover, despite the pronounced heterogeneity among the three datasets, OKR-CELL maintains robust and superior performance in all settings, establishing a new SOTA for single-cell foundation models.

\subsubsection{Zero-shot and Few-shot Cell Type Annotation}

\paragraph{\textbf{Few-shot Classification.}}
We evaluated the model's adaptability in data-scarce scenarios by varying the number of support samples ($K\ \in\{1,\ 5,\ 9\}$) per class for model finetuning. In the few-shot classification setting \cite{snell2017prototypical}, all employed models add a linear layer as the classification head and freeze the backbone part during being finetuned. As illustrated in Figure \ref{fig:2}D, OKR-CELL consistently outperforms baseline methods across all settings on both datasets. Notably, with 9-shot supervision on the Zheng68k \cite{zheng2017massively} dataset, our model achieves a peak Accuracy of 0.787, establishing a significant margin over scCLIP-GPT (0.554) and LangCell (0.190). A similar trend is observed in the Great Apes dataset, where OKR-CELL reaches an Accuracy of 0.412, compared to 0.361 for LangCell and 0.264 for scCLIP-GPT. These results indicate that OKR-CELL possesses superior data efficiency, effectively extracting discriminative features even with minimal annotated samples.

\paragraph{\textbf{Zero-shot Classification.}}
To assess model generalization to unseen tissues, we conducted zero-shot classification \cite{socher2013zero,xian2017zero} on the Eye, Prostate Gland \cite{joseph2021single}, and Great Apes \cite{jorstad2023comparative} datasets. As shown in Figure \ref{fig:2}E, OKR-CELL demonstrates robust quantitative performance, particularly on the challenging Prostate Gland dataset, where it achieves an Accuracy of 0.511, substantially surpassing the runner-up scCLIP-GPT (0.299). This advantage is qualitatively corroborated by the Sankey diagrams in Figure \ref{fig:2}F while LangCell exhibits chaotic and entangled misclassifications, OKR-CELL displays coherent mapping flows between predicted and ground-truth labels. Specifically, OKR-CELL can correctly classify most of basal cells of prostate epithelium, even if being misclassified, the cells are assigned to relatively related cell types, such as epithelial cells of the urethra. By contrast, the classification accuracy of LangCell for basal cells of prostate epithelium is less than half, and it produces more unreasonable misclassifications. These results further confirm the advance of our OKR-CELL model to align cell representation with biological semantics without additional prior supervision. 

\begin{figure*}[!tpb]
    \begin{center}
\includegraphics[width=0.98\linewidth]{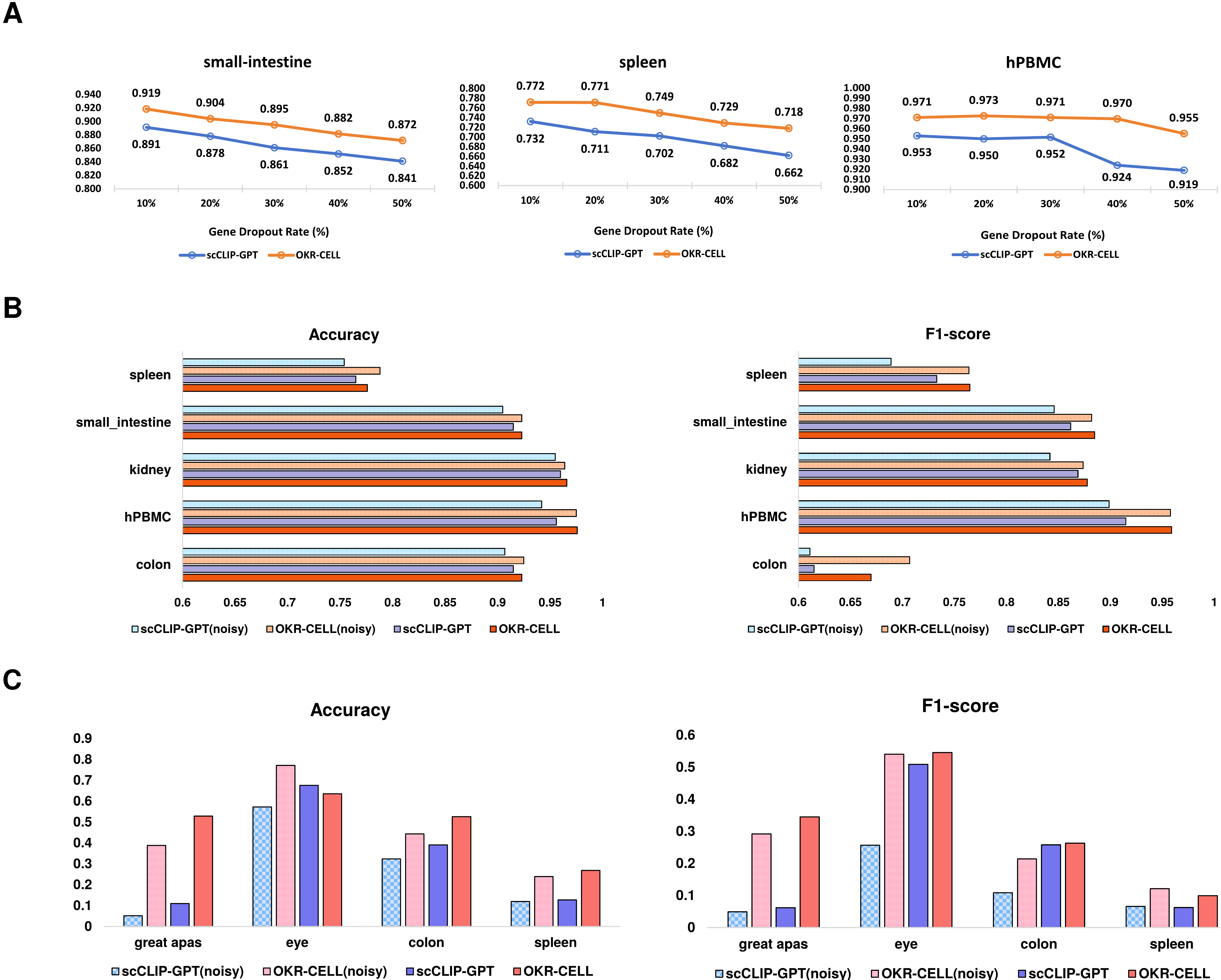}
    \end{center}
    \caption{(A) Curve plot of cell type annotation performance of scCLIP-GPT and OKR-CELL under varying gene dropout rates in input scRNA-seq data. (B) Bar plot of traditional cell type annotation performance by scCLIP-GPT and OKR-CELL trained under noisy cell-text data, where the obtained methods denoted by scCLIP-GPT(noisy) and OKR-CELL(noisy). (C) Bar plot of zero-shot cell type annotation performance by scCLIP-GPT(noisy) and OKR-CELL(noisy) along with original scCLIP-GPT and OKR-CELL. 
    }
    \label{fig:4}
\end{figure*}



\subsection{Results under Noisy Data}
In this part, we investigate in how noise in data affects the methods. First, to testify the robustness of different models to noise in test data, we randomly knocked out fragments of gene transcription sequences in the test data to simulate the negative impacts on data quality caused by various factors during gene sequencing. And we validate its impact on cell type annotation task. Subsequently, we incorporated noise into the multi-modal data. Specifically, we uniformly add noise to the training data by tampering with the gene expression values of a certain proportion of genes; to mimic the noise induced by mismatches or partial matches between cell-text data, we shuffle the pairing relationships between cells and text in a subset of the paired data. To evaluate the impact of multi-modal noise, we conducted tests in both traditional (intra-modal) and zero-shot (cross-modal) cell type annotation tasks. In the experiments, we employ our baseline model scCLIP-GPT and our OKR-CELL for performance comparison.

\subsubsection{Results under Corrupted Testing Data}
Figure \ref{fig:4}A illustrates the cell type annotation performance of scCLIP-GPT and OKR-CELL under increasing gene dropout rates (10\%–50\%) in input scRNA-seq data, tested on small-intestine, spleen, and hPBMC datasets. On small-intestine dataset, when the dropout rate increases from 10\% to 50\%, OKR-CELL still outperforms scCLIP-GPT. Concretely, the percentage decrease in OKR-CELL’s accuracy 5.1\% (91.9\% $\rightarrow$ 87.2\%) is lower than that of scCLIP-GPT 5.6\% (89.1\% $\rightarrow$ 84.1\%). For spleen dataset, when gene dropout rate grows from 10\% to 20\%, OKR-CELL’s performance barely degrades, while scCLIP-GPT's performance drops by 2.86\%. Especially on hPBMC dataset, with the dropout rate increasing from 10\% to 40\%, OKR-CELL’s accuracy remains above 97\%, while scCLIP-GPT’s plummets from 95.2\% to 92.4\%. These results confirm OKR-CELL’s strong tolerance to gene expression data loss, a critical advantage in real-world scRNA-seq experiments where incomplete gene coverage is common.

\subsubsection{Results under Noisy Mulit-modal Training Data}
To simulate the noise in multi-modal data, we set the shuffling percentage of gene expression values and pairwise cells-text relationship to 30\%. 

\paragraph{\textbf{Traditional Cell Type Annotation.}} 
First, we compared the performance of scCLIP-GPT and OKR-CELL trained on noisy data in traditional cell annotation tasks, and we also included the results of models trained on original data for comparison. As shown in the Figure \ref{fig:4}B, the performance of scCLIP-GPT trained on noisy data degrades on most datasets, with the F1-score dropping by 6\% (from 73.3 to 68.9) on the spleen dataset; in contrast, our OKR-CELL shows no significant performance decline across all datasets and outperforms the former significantly. For example, on the colon dataset, the F1-score of OKR-CELL(noisy) is 15.7\% higher than that of scCLIP-GPT(noisy), which is notably larger than the gap between the two under non-noisy conditions.
Surprisingly, OKR-CELL(noisy) trained on fabricated data even achieves performance improvements on some datasets. For instance, the F1-score for classification on the colon dataset increases by 5.5\% (from 67.0 to 70.7), and the accuracy on the spleen dataset rises by 5.5\% (from 77.6 to 78.8). We analyze the potential reasons for this result as follows: 1) The cell encoder is not very sensitive to the expression values of input transcriptome genes, and the noise injection method through expression value perturbation actually enhances the robustness of the cell encoder; 2) The cell types in the colon dataset have a large number of similar or identical cell samples in the original training set, which helps the model learn strong discriminative capabilities for such cells from noisy data. Overall, our OKR-CELL model exhibits excellent robustness to noisy training data, which enables it to achieve better performance in more extreme data scenarios.

\paragraph{\textbf{Zero-shot Cell Type Annotation}}
Furthermore, we evaluated the performance of scCLIP-GPT and OKR-CELL in zero-shot cell classification tasks across several challenging datasets and their noisy counterparts. As illustrated in the Figure \ref{fig:4}C, the zero-shot performance of scCLIP-GPT trained on noisy data experiences significant fluctuations or severe degradation; specifically, on the great apas dataset, the F1-score of scCLIP-GPT(noisy) drops to a mere 4.9\% (from 6.16\%). In comparison, our OKR-CELL demonstrates superior robustness in this zero-shot scenario, substantially outperforming the former across all metrics. Notably, on the eye dataset, the F1-score of OKR-CELL(noisy) reaches 54.06\%, which is 28.37\% higher than that of scCLIP-GPT(noisy), further widening the gap under noisy conditions. Remarkably, OKR-CELL(noisy) even achieves performance improvements on certain datasets; for instance, the accuracy on the eye dataset rises by 13.62\% (from 63.53\% to 77.15\%), and the F1-score on the spleen dataset increases from 9.91\% to 12.12\%. We analyze the potential reasons for this result as follows: 1) The noise injection acts as a form of data augmentation that effectively prevents the model from overfitting to complex local distributions, thereby improving generalization; 2) The architecture of OKR-CELL possesses a stronger capability to capture key biological features, enabling it to identify more robust cellular identities from perturbed expression profiles. In summary, these results underscore OKR-CELL’s inherent advantage in learning generalizable cell-text semantic representations: its performance remains robust whether trained on clean or noisy cell-text data, making it highly suitable for zero-shot cell type annotation tasks where labeled data is scarce.

\begin{figure*}[h!]
    \begin{center}
\includegraphics[width=0.95\linewidth,height=15cm]{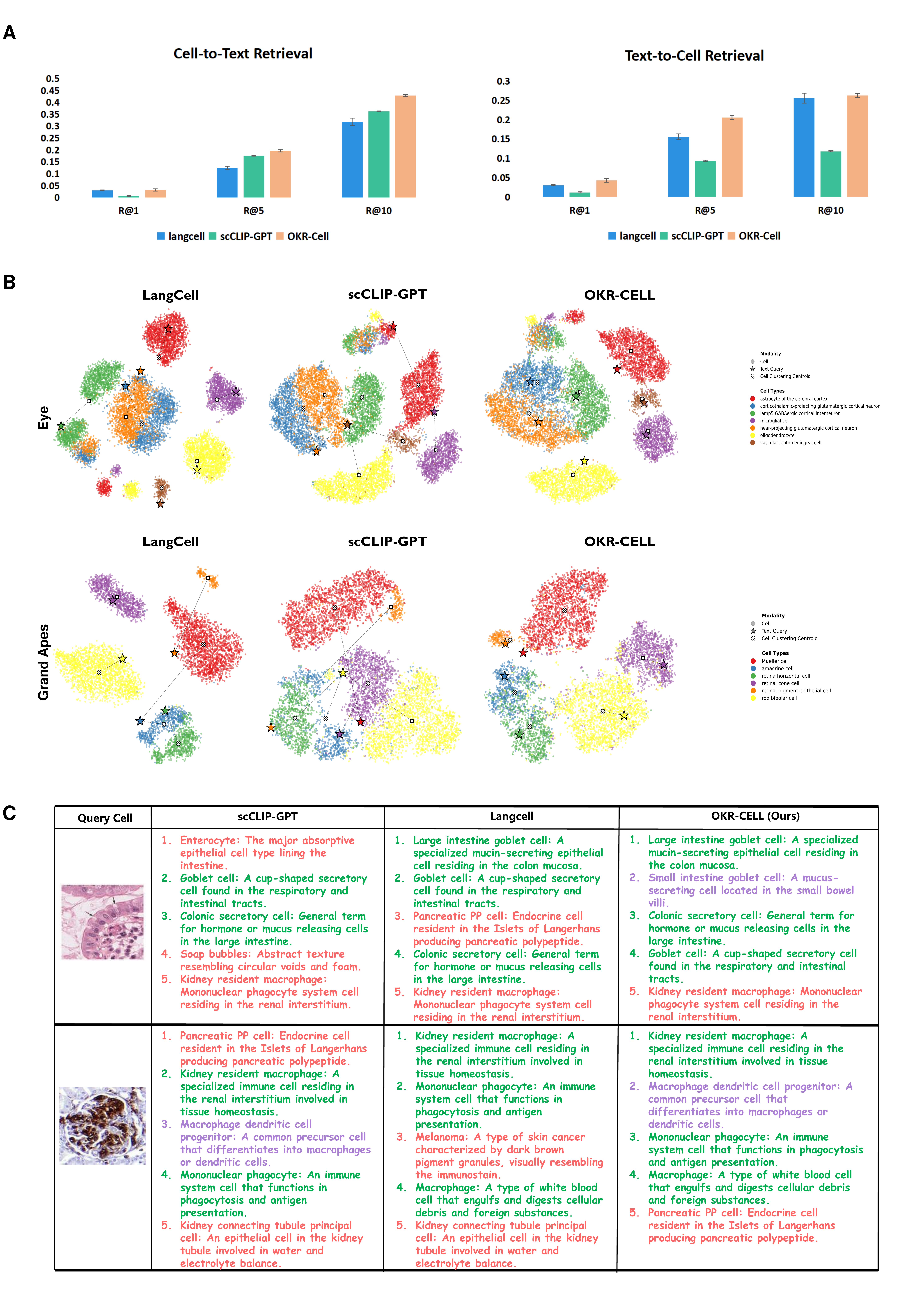}
    \end{center}
    \caption{(A) Bar plot of bidirectional cross-modal retrieval performance (R@1,5,10) by different methods on our proposed SCxGEN-CT5K dataset. (B) T-SNE visualization plots of cell embeddings and textual embeddings of their corresponding cell types generated by different methods on eye and Great Apes reference datasets colored by ground-truth cell types. (C) Results visualization of cell based text retrieval on SCxGEN-CT5K dataset. The ground-truth and non ground-truth descriptions are marked in green and red, respectively. Note that, in addition to the original cell’s paired cell-type textual description, any text descriptions corresponding to other subtypes under the same parent cell type are also marked in green. Samples with correct broad-type but incorrect subtype are considered partially correct and are highlighted in purple.
    }
    \label{fig:3}
\end{figure*}

\subsection{Cross-modal Retrieval for Novel Cell Types}
To further testify the cross-modal representation ability of our method, we conduct bi-directional cross-modal retrieval experiments, i.e. cell-based text search and text-based cell search. First, we establish a testing dataset named SCxGEN-CT5K, which contains 5K cell-text pairs from 36 tissues randomly selected from CELLxGENE platform, which are all unseen in our training data. To assess cross-modal retrieval performance, we split SCxGEN-CT5K into 5 equal folds and report the average recall rate (R@k) (details in Section \ref{sec_r@k}) over these 5 folds. Note that for the recall metric computation, we consider only those samples with strictly one-to-one annotated cell–text pairs as positive; all other samples including those sharing the same cell type or subtype are treated as negatives.  

From Figure \ref{fig:3}(A), the experiments evaluate Cell-to-Text and Text-to-Text Retrieval using R@1, R@5, R@10 metrics. For Cell-to-Text Retrieval, OKR-CELL achieves 3.32 on R@1, outperforming LangCell by 2.1\% and scCLIP-GPT (~0.67). For R@5 and R@10, OKR-CELL arrives at 19.7 and 42.9, achieving absolute boost of (11.9\%, 16.0\%) compared to second best scCLIP-GPT. For Text-to-Cell Retrieval, our method also outperforms other two competitors. These results confirming superiority advance of our method. Moreover, to better analysis the cross-modal joint embedding space, we utilize t-SNE to visualize the learned cellular and textual representations of labels from eye and Grand Apes dataset. See Figure \ref{fig:3}(B), we can find our OKR-CELL is able to associate the cells and their corresponding classes. For example, on eye dataset, only our method can accurately assign the label embeddings marked in \textcolor{blue}{blue} and \textcolor{orange}{orange} color to the centroid of cells belonging to them, respectively. On Great Apes dataset, our method is the only one that can accurately establish the category mapping attributable to each cell cluster. In Figure \ref{fig:3}(C), we select two cases of cell-based text search and list top-5 most relevant texts. For goblet cell-related queries, OKR-CELL prioritizes tissue-specific subtypes (large/small intestine goblet cells) as top results, ensuring functional coherence. scCLIP-GPT misranks "Enterocyte" (biologically irrelevant), while LangCell includes unrelated "Pancreatic PP cell". For kidney resident macrophage queries, OKR-CELL clusters lineage-related cells (progenitors, mononuclear phagocytes) logically, whereas LangCell incorrectly lists "Melanoma" and scCLIP-GPT fails to prioritize lineage relationships.

\section{Discussion}\label{sec12}

In this paper, we present OKR-CELL, a robust single-cell foundation model grounded in a cross-modal cell–language pretraining framework and augmented with open-world biological knowledge. To overcome two critical limitations of current Cell-Text Pretraing (CTP) based approaches: shallow integration of cellular context and inattention of noise in multi-modal data, OKR-CELL synergistically combines large language model (LLM) enhanced textual corpora with a novel Cross-modal Robust Alignment (CRA) objective for cross-modal cell-text pretraining. Pretrained on 32 million cell–text pairs, OKR-CELL achieves state-of-the-art performance across six benchmark tasks: unsupervised cell clustering, cell type annotation under conventional, few-shot, and zero-shot settings, batch effect correction, and bidirectional cell–text retrieval. These results underscore its exceptional robustness, generalization capability, and potential as a foundational model for multi-modal single-cell analysis. 

Our method achieves impressive performance across a broad spectrum of single-cell analysis tasks. On conventional benchmarks, including unsupervised cell clustering and supervised cell type annotation, it consistently delivers strong results. Considering our primary contribution aiming at enhancing model generalization, we first evaluate OKR-CELL on few-shot cell type annotation. Experimental results show that OKR-CELL substantially outperforms two comparison methods, LangCell and scCLIP-GPT, by significant margins (e.g., +42.1\% in accuracy on Zheng68k dataset for 9-shot classification).

Moreover, leveraging cross-modal cell–text similarity, our model demonstrates the ability to recognize cell types never seen during training, addressing the zero-shot cell annotation task as a coarse-grained form of cross-modal alignment. In this setting, OKR-CELL also ourperforms LangCell by a large margin (e.g., +70.9\% in accuracy on Prostate Gland dataset). To further assess fine-grained cross-modal understanding, we introduce SCxGEN-CT5K dataset, a novel cell-text dataset designed for bidirectional cell–text retrieval under open-set setting. On this challenging benchmark, OKR-CELL achieves the best reported results, highlighting its capacity to capture nuanced semantic correspondences between cellular states and natural language descriptions.

Furthermore, to validate the robustness of our approach against multi-modal noise, a common issue in real-world single-cell data, we deliberately inject two types of perturbations during pretraining: (1) random perturbations to gene expression values, and (2) synthetic mismatches between cells and their associated textual descriptions. Under these noisy conditions, OKR-CELL maintains superior performance compared to scCLIP-GPT, with particularly pronounced gains in zero-shot annotation tasks that rely heavily on cross-modal alignment. Remarkably, on certain datasets (e.g., eye and spleen dataset), models trained on our corrupted data even outperform those trained on original data. This success suggests that our approach can confers enough robustness, which likely stems from our model can distinguish the unreliable false positive sample pairs and enhance the discriminability between negative pairs. 

Note that the two core innovations of our method are orthogonal to those employed in current mainstream single-cell foundation models. As such, they do not conflict with existing architectures; rather, they can be seamlessly incorporated into them. In this work, we adopt scGPT as the backbone for the cellular encoder, and demonstrate that OKR-CELL, built upon this architecture, achieves substantial performance improvements over the scGPT backbone as baseline model. We further anticipate that our contributions can be effectively integrated with a wide range of other single-cell foundation models including CTP based methods (\textit{e.g.} LangCell \cite{zhao2024langcell}), offering complementary enhancements in representation learning and robustness.

In the future, we plan to extend our model to integrate multi-omics data, leveraging textual information as a unifying interface to enable deeper, more coherent interpretation of heterogeneous molecular layers. In the present work, we employ cell-level textual descriptions to enhance global cellular representations. In future work, we aim to incorporate gene-level textual annotations as fine-grained, local supervisory signals to strengthen the model’s capacity for capturing nuanced, locus-specific biological semantics. Furthermore, we intend to build upon OKR-CELL to develop a chat-based multi-modal Large Language Models \cite{liu2023visual,chen2024internvl,wang2025sega} capable of detailed, interactive interrogation of single-cell data. Unlike existing approaches such as CellWhisperer \cite{schaefer2025multimodal} that aligns bulk-level transcriptomic profiles with text, we propose to perform cross-modal alignment directly at the single-cell resolution, using our LLM-augmented, cell-specific transcriptome–text corpora to achieve finer-grained cell understanding. Ultimately, we hope that our contributions, both in terms of high-quality multi-modal data curation and robust, cross-modal aligning methodology, will advance the development of single-cell foundation models and catalyze progress in downstream applications, thereby expanding the frontier of our understanding and interpretability of cellular biology.

\section{Methods}\label{sec3}

In this section, we provide a detailed description of the OKR-CELL method, including data curation, model architecture, and training objectives.

\subsection{Single-cell RNA-seq encoder}\label{sec31}
To model the transcriptomic profiles of single cells, we adopt a transformer-based architecture derived from scGPT as the cellular encoder, with detailed implementations and functional designs elaborated below.

\subsubsection{Input Embeddings}\label{sec311}
The input embedding module transforms single-cell RNA-seq data—structured as a cell-gene matrix \(\mathbf{X} \in \mathbb{R}^{N \times G}\), where $N$ denotes the total number of cells, $G$ is the total count of genes, and \(X_{i,j}\) represents the raw read count of gene \(j\)in cell \(i\)) into a consistent latent representation through three interconnected modules, followed by feature fusion:

\textbf{1. Gene Tokenization.}
Each gene is treated as a basic information unit (analogous to a word in NLG) and assigned a unique integer ID \(id(g_j)\). Special tokens (e.g., $<$cls$>$ for cell representation aggregation, $<$pad$>$ for input length padding) are included to support cross-study gene set integration. For cell \(i\), the gene token sequence is defined as:  

Each gene is treated as a fundamental informational element (analogous to a token in natural language processing) and assigned a unique integer identifier \(id(g_j)\). To support cross-study integration of gene sets, two special tokens are introduced: $<$cls$>$ for aggregating cell-level representations and $<$pad$>$ for padding sequences to a uniform length. For cell \(i\), the gene token sequence is defined as:

\begin{equation}
t_{g}^{(i)} = \left[id(g_{1}^{(i)}), id(g_{2}^{(i)}), \ldots, id(g_{M}^{(i)})\right]
\label{eq:gentok}
\end{equation}
where \textit{M} (predefined input length) is set to the number of selected highly variable genes (HVGs), a standard practice in single-cell transcriptomic analysis.

\textbf{2. Gene Expression Binning.}
To resolve scale inconsistencies across different sequencing batches—a challenge that cannot be fully addressed by TPM normalization or log1p transformation alone, we implement a value binning strategy with standardized parameters:

1). Preprocessing: Log1p transformation and HVG selection are performed first;   

2). Binning Operation: Non-zero expression values of each cell are partitioned into \(O\) equal-interval bins, while zero values are retained as 0. The binned value \(x_j^{(i)}\) is formally defined as:  

\begin{equation}
\begin{aligned}
x_{j}^{(i)} = \begin{cases} 
k, & \text{if } X_{i,j} > 0 \text{ and } X_{i,j} \in [o_k, o_{k+1}] \\
0, & \text{if } X_{i,j} = 0 
\end{cases}
\end{aligned}
\label{eq:genexp}
\end{equation}
The final expression vector for cell \(i\) is \(x_{e}^{(i)} = [x_{1}^{(i)}, x_{2}^{(i)}, ..., x_{M}^{(i)}]\), ensuring consistent semantics of expression levels across batches.

\textbf{3. Embedding Fusion.}
Gene tokens and binned expression values are independently projected to \(D\)-dimensional vectors via embedding layers (\(emb_g\), \(emb_x\)) respectively. The final input embedding for cell \(i\): is obtained by element-wise summation of these two components, integrating both gene identity and expression magnitude information:

\begin{equation}
\begin{aligned}
h^{(i)} = emb_g(t_{g}^{(i)}) + emb_e(x^{(i)})
\end{aligned}
\label{eq:embfus}
\end{equation}
where \(h^{(i)} \in \mathbb{R}^{M \times D}\) serves as the input to subsequent Transformer encoder blocks.

\subsubsection{Model Architecture and Inference}\label{sec312}

The cellular encoder retains scGPT’s self-attention Transformer backbone to capture complex gene-gene interaction patterns embedded in the input embedding \(\mathbf{h}^{(i)}\). The encoding process involves feeding \(\mathbf{h}^{(i)} \in \mathbb{R}^{M \times D}\) through a stack of Transformer blocks, which can be formulated as:
\begin{equation}
\begin{aligned}
h_0^{(i)} = h^{(i)}, 
\quad h_l^{(i)} = \text{transformer\_block}(h_{l-1}^{(i)}) \quad \forall l \in [1, n]
\end{aligned}
\end{equation}

where \textit{n} denotes the number of Transformer blocks, and the final output \(h_n^{(i)} \in \mathbb{R}^{M×D}\) provides a comprehensive representation that supports both gene-level and cell-level downstream tasks.

\textbf{Cell-level Representation.} We adopt a special $<$cls$>$ token at the beginning of genes tokens, which enables the model to learn task-adaptive pooling within Transformer blocks. After processing through \textit{n} Transformer blocks, the embedding corresponding to the $<$cls$>$ token in the final output \(h_n^{(i)}\) is extracted as the cell representation $h_{cls}^{(i)} \in \mathbb{R}^D$, which aggregates global information of the cell’s transcriptomic profile.

\textbf{Attention Masking Strategy.} We employ the masked attention strategy in scGPT \cite{cui2024scgpt} during self-attention computation. We first divide the input tokens into three groups:
\begin{itemize}
\item $<$cls$>$ token: For cell embedding aggregation.
\item Known genes: Genes with seen token/expression embeddings.
\item Unknown genes: Targets for expression value prediction (randomly sampled during training).
\end{itemize}
The principle of mask generation is ``predicting the expression value of Unknown genes based on known genes''. The binary attention mask (\(M \times M\), matching the input token length \textit{M}) restricts attention computation to "known" tokens when predicting "unknown" ones. 

\textbf{Inference Manner.} The inference of the model is performed by two iterative steps: \textbf{Gene-prompt Generation} and \textbf{Cell-prompt Generation}. First, gene-prompt generation aims to predict the expression values of unknown genes using a subset of ``known gene'' as the initial prompt, meanwhile outputs a $<$cls$>$ token embedding. Afterwards, for cell-prompt generation, the model take the obtained $<$cls$>$ token embedding as input $<$cls$>$ tokens for incorporating cell type into model as context. The forward propagation process similar to gene-prompt generation is repeated. Finally, we obtain the cell representation $h_{cls}^{(i)}$.


\subsection{Training Data Curation}\label{sec_Data}
In this section, we introduce how to construct the cell-text dataset for our model pre-training. We first collect 32 million single-cell data from the CellxGene platform \cite{czi2025cz}, including scRNA-seq matrices and corresponding metadata annotations from diverse human
tissues and organs. The textual descriptions of each cell are generated with
cellular metadata and the OBO Foundry \cite{smith2007obo}. In the next part, we illustrate how to leverage LLM to incorporate open-world knowledge into the textual descriptions of cells in detail. 

\subsubsection{LLM-enriched Textual Corpus Curation}
The most intuitive plan to introduce open-world knowledge is directly employing LLM to rewrite original textual descriptions. However, considering the scarcity and rigor of specialized expertise in the biomedical domain, we anticipate that this approach will inevitably introduce a significant amount of hallucinated information. Therefore, we additionally integrate RAG \cite{lewis2020retrieval,gao2023retrieval} (Retrieval-Augmented Generation) and Reliability Screening operation into this pipeline. The core workflow is operated as following four steps (see Figure \ref{fig:0}A): 

\textbf{(1) Data Preprocessing and Structuralization.} Initially, a standardization process is applied to the terminology and metadata annotations within the textual descriptions. Specifically, information including disease, gender, tissue and cell type is mapped to corresponding standardized textual descriptions by referencing the Cell Ontology database.

\textbf{(2) Constructing a Biomedical Knowledge Base.} This step includes thee sub-steps:  

\textbf{- Data Source Integration}: We aggregate domain-specific resources (e.g., papers from PubMed database) to form a comprehensive biomedical knowledge corpus. Specifically, we only extract the abstract part of each paper for condensing knowledge. 

\textbf{- Text Cleaning \& Chunking}: The corpus is processed to remove noise (e.g., redundant formatting) and split into small, semantically coherent snippets (e.g., 500-token chunks) to optimize retrieval relevance.

\textbf{- Vector Database Construction}: Each cleaned snippet is encoded into a dense vector using a biomedical text encoder (e.g., BioBERT \cite{lee2020biobert}). These vectors are stored in a vector database to enable efficient semantic similarity search.

\textbf{(3) RAG enhanced Textual Descriptions Generation.}

For RAG, we first split the preprocessed input text and obtain the chunked information including (\texttt{Cell Type, Tissue, Sex, Development Stage}), followed by combining them into a templated query text. Then, the BioBERT encoded this query text into feature and used it to search the vector database, retrieving the top-$K$ most semantically similar knowledge snippets $(K_{i})$.

Afterwards, we select the Deepseek-V3 \cite{liu2024deepseek}, an cutting-edge LLM for text augmentation. Specifically, to generate augmented text, we inject both the standardized original text (from Step (1)) and the retrieved knowledge snippets into our designed prompt template, followed by feeding the combined prompt sequence into the LLM. Our prompt template is elaborately crafted to incorporate more biomedical knowledge into cellular text descriptions, whose details can be seen in the Section \ref{Supp_PromptTemp} of supplementary materials. Benefiting from the enormous volume of information contained in LLM (and the targeted domain knowledge from RAG), the generated text can not only retain the key concepts and semantic information of the original text, but also enriches more related biomedical knowledge to assist understanding a cell from more comprehensive perspectives.

\textbf{(4) Reliability Screening.} Despite LLM has remarkable capabilities, hallucination problem remains unresolved. A key issue in LLM-based text augmentation is the enriched texts may fabricate content that deviates from the original descriptions. In result, these concocted contents will have a negative impact during model training.

To mitigate these hallucinations, we integrates a Reliability Screening (RS) operation to exclude unfaithful augmented texts. Concretely, Leveraging the Clinical-LongFormer as textual feature extractor $f_{RS}(\cdot)$, RS operates in two steps: 1) Encode original text \(T_{i}^{ori}\) and augmented text \(T_{i}^{aug}\) into dense vectors via Clinical-LongFormer; 2) Calculate their semantic relevance using cosine similarity: \(s(T_{i}^{ori},T_{i}^{aug})=\frac {f_{RS}(T_{i}^{ori})^{\top }\cdot f_{st}(T_{i}^{aug})}{\left\| f_{st}(T_{i}^{ori})\right\| \left\| f_{st}(T_{i}^{aug}\right) \| }\).   

A predefined threshold \(\alpha\) determines whether the augmented text is qualified: \(s < \alpha\) means the augmented text is discarded, while \(s \geq \alpha\) means it is retained as semantically faithful. Consequently, RS effectively reduces noise in the augmented texts and uphold the training data quality.

\subsection{Learning Objective}
The pre-training objective of our model can be divided into two categories: intra-modal learning and cross-modal learning. The former is designed to enhance representation ability of model by performing generative pre-training on cellular RNA-seq data, the latter aims to transfer rich knowledge from textual descriptions into cellular domain via cross-modal aligning. The details of them are depicted as following parts. 

\subsubsection{Intra-modal Cellular Generative Pre-training}
\label{ICGP}
To learn biologically discriminative representations of cells and genes, we follow scGPT to incorporate two tailored self-supervised objectives: Gene Expression Prediction (GEP), Gene Expression Prediction for Cell Modelling (GEPC). 

\textbf{Gene Expression Prediction (GEP)} aims to capture gene-gene interactions via self-supervised learning. For each cell, a random subset of genes is masked (with the mask ratio randomly selected from \{0.25, 0.50, 0.75\} per training iteration, and a 2-layer multi-layer perceptron (MLP) is used to infer the binned expression values of the masked genes from the final Transformer output \(\mathbf{h}_n^{(i)}\). The objective minimizes the cross-entropy (CE) loss exclusively over the masked positions \(\mathcal{M}_{mask}\), ensuring the model learns meaningful dependencies between genes:
\begin{equation}  
\begin{aligned}   
\tilde{x}^{(i)} &= MLP\left(h_{n}^{(i)}\right), \\
\mathcal{L}_{GEP} &= \frac{1}{|\mathcal{M}_{mask}|} \sum_{j \in \mathcal{M}_{mask}} ce\left(\tilde{x}_{j}^{(i)}, x_{j}^{(i)}\right).
\end{aligned}
\end{equation}


\textbf{Gene Expression Prediction for Cell Modelling (GEPC)} explicitly enhances cell representation learning by predicting gene expressions from the cell embedding \(h_c^{(i)}\). For each gene \textit{j}, a query vector \(q_j\) is derived from its token embedding, and the predicted expression is computed via a parameterized inner product between \(q_j\) and the cell embedding \(\mathbf{h}_c^{(i)}\). The loss is defined as:
\begin{equation}  
\begin{aligned}
q_j &= MLP\left(emb_g\left(t_g^{(i)}\right)\right), \\
\tilde{x}_j^{(i)} &= q_j \cdot W \cdot h_c^{(i)}, \\
\mathcal{L}_{GEPC} &= \frac{1}{|\mathcal{M}_{mask}|} \sum_{j \in \mathcal{M}_{mask}} ce\left(\tilde{x}_j^{(i)}, x_j^{(i)}\right).
\end{aligned}
\end{equation}
where \(W \in \mathbb{R}^{d \times d}\) is a learnable weight matrix that bridges gene query vectors and cell embeddings. GEPC inherits gene token embeddings $emb_g(t_{g}^{(i)}) $ from Eq. \ref{eq:embfus}. Combining GEP and GEPC yields synergistic performance gains: GEP captures local gene-gene dependencies, while GEPC enforces global consistency between cell-level characteristics and gene expression profiles.


\subsubsection{Cross-modal Cell-text Pre-training}
\label{CRA}
To make our cross-modal alignment objective robust to the noise contained in the cell-text pairs, we propose a novel learning objective named Cross-modal Robust Alignment (CRA) objective, which is composed of three components: Progressive Sample Weighting (PSW) strategy, Coupled Momentum-updated Based Memory Bank (CMMB), and Cross-modal Complementary Alignment (CCA) loss. On the whole, we first substantially increases the diversity of negative samples within the batch through constructing CMMB. Then, we integrate obtained MMB into our CCA loss to further enhance its generalization ability. Lastly, we apply PSW strategy to dynamically modify the importance of individual samples according to their properties.  

\paragraph{\textbf{Coupled Momentum-updated Memory Bank (CMMB)}} 
In order to enrich the negative sample diversity and scale for both cellular and textual modalities, we follow \cite{he2020momentum,wang2022coder} to adopt the momentum update to build two coupled dynamic memory banks \(\mathbf{B}_c\) and \(\mathbf{B}_t\), which are dedicated to preserving more cellular and textual features. The weights of these momentum encoders are adjusted dynamically based on their modality-specific encoders through momentum updates. For each latest training iteration, visual and textual instances are processed by the momentum encoders to generate corresponding embeddings, and these features are then stored in the coupled memory banks, which are implemented by utilizing the data structure of queue. The size of memory banks is represented by \textit{B}.

\paragraph{\textbf{Progressive Sample Weighting (PSW) Strategy} }

It is well known that for model optimizing, if a model is exposed to hard samples too early before it has acquired basic discriminative capabilities, it may suffer from unstable gradients, or even premature convergence to suboptimal solutions. Conversely, if hard samples are consistently under-weighted or ignored, the model may fail to learn critical decision boundaries, leading to poor robustness on challenging cases. To address this dilemma, inspired by the pedagogical principle of curriculum learning (CL) \cite{bengio2009curriculum,wang2021survey}, which posits that learning is most effective when structured from simple to complex concepts. We propose a novel adaptive sample reweighting mechanism that dynamically adjusts the importance of training samples based on their instantaneous difficulty and the current stage of model training. Unlike static weighting schemes that treat all samples uniformly throughout training, our approach explicitly encodes the “easy-to-hard” learning trajectory into the optimization process, enabling more stable convergence and improved generalization. 

To achieve it, we design a two-phase adaptive weighting strategy: \textbf{(1) Early Training Phase}: Prioritize easy samples to establish a reliable foundation of feature representation and decision logic; \textbf{(2) Late Training Phase}: Shift focus toward hard samples to refine model behavior on ambiguous or high-stiffness instances, thereby enhancing robustness and generalization. This mirrors how humans acquire skills—first mastering fundamentals before tackling advanced problems and aligns with empirical observations in deep learning that early-stage training benefits from low-variance signal sources.

Let $\mathcal{L}_{pos}(C_i, T_j)$ denotes the loss of the cross-modal similarity of negative sample pairs $(C_i, T_j)$. The difficulty of this pair is quantified by this loss value: higher loss implies greater difficulty. To implement the curriculum schedule, we introduce a dynamic difficulty threshold $\gamma(e)$, which evolves smoothly over the course of training. The adaptively based on a phase-switching strategy governed by a schedule indicator $\gamma(e)$ is computed via a power-law schedule:
\begin{equation}  
\begin{aligned}
\gamma(e) = \gamma_{\text{start}} + (\gamma_{\text{end}} - \gamma_{\text{start}}) \cdot e^{\alpha}, \ 
e = \frac{t}{T} \in [0,1]
\end{aligned}
\end{equation}
where $\gamma_{\text{start}}=2$ and $\gamma_{\text{end}}=10$
 are the initial and final values of the threshold, respectively, and $\alpha=0.5$ controls the pace of transition. The current global iteration and total iterations of training are denoted by $t$ and $T$. 

Then, the weight of sample pairs weight $w^{(C_i, T_j)}$ is assigned adaptively based on a phase-switching strategy governed by the schedule indicator $\gamma(p)$ :
\begin{equation}  
\begin{aligned}
w^{PSW}{(C_i, T_j)} = 
\begin{cases} 
\max\left(0,\ 1 - \frac{\mathcal{L}_{\text{neg}}^{(C_i, T_j)}}{\gamma(e)}\right), & \text{if} \ e < 0.2 \, \\
\max\left(0,\ 1 + \frac{\mathcal{L}_{\text{neg}}^{(C_i, T_j)}}{\gamma(e)}\right), & \text{otherwise} 
\end{cases}
\end{aligned}
\end{equation}
In the early training phase ($p < 0.2$), the model prioritizes easy samples by assigning them with higher weights, while suppressing hard samples whose loss exceeds the threshold. As training progresses into the later phase, the weighting scheme flips: hard samples receive amplified weights, enabling the model to focus on refining its understanding of challenging cases—effectively realizing the “easy-to-hard” curriculum paradigm in a continuous and adaptive manner.

\paragraph{\textbf{Cross-modal Complementary Alignment (CCA) Loss}} Traditional Cross-Modal Contrastive Learning (CMCL) is calculated based on sample pairs, which maximizes the mutual information between positive pairs and minimize those of negative pairs. However, due to the noise implying in dataset, such as the limitations of gene sequencing technologies and the cellular understanding reflected in textual descriptions, there exist some "false positive sample pairs" or "noisy positive sample pairs" \cite{lan2023towards,qin2023cross}. Consequently, these supervision signals of positive samples will mislead the model by CMCL, leading to performance degradation via learning erroneous correlations.

To tackle this problem, distinct processing approaches are applied to handle positive and negative sample pairs respectively. For positive pairs, we introduce a reliability indicator employed as a soft label to quantify the credibility of positive sample pairs. Concretely, we evaluate the reliability of positive sample pairs from two perspectives:
(1) \textbf{Symmetry}: A positive sample pair is deemed more reliable only when the cross-modal similarity between the samples within the pair is high for both directions of matching. Since the credibility of positive samples is ensured only if the cross-modal similarities calculated from any modality to another one are both high. 
(2) \textbf{Stability}: From a temporal perspective, a positive sample pair is considered credible only if the degree of change in its cross-modal similarity across different training epochs is relatively stable.



From perspective of symmetry, the reliability indicator about cross-modal symmetry is defined as:
\begin{equation}  
\begin{aligned}
Rel^{sym}_{i,i} &= \frac{1}{2}\left(
    \frac{\exp(S_{i,i}/\sigma)}{ \sum_{k=1}^N \exp(S_{i,k}^{C \to T}/\sigma)} + 
    \frac{\exp(S_{i,i}/\sigma)}{ \sum_{k=1}^N \exp(S_{i,k}^{T \to C}/\sigma)} \right) 
\end{aligned}
\label{W_s_pos}
\end{equation}
where $\sigma$ is a temperature parameter. 

Then we illustrate how to measure the temporal stability of positive pairs. We first employ a memory container $\mathbf{Q}$ to record individual reliability indicator $Rel^{sym}$ of each positive pairs per epoch. Specifically, $\mathbf{Q} \in \mathbb{R}^{M \times P}$ is a look-up table, where $M$ is the number of training samples and $P$ is the number of maximum training epochs. The historical training behavior of pair $(C_i,T_i)$ by taking the standard deviation of $Rel^{sym}_{i,i}$ before $p$-th epoch ($p > 1$):
\begin{equation}  
\begin{aligned}
\mu_{i,i}(p) &= \frac{1}{P-1} \sum_{p=1}^{P-1} Rel^{sym}_{i,i}(p), \quad 
q_{i,i}(p) = \sqrt{\frac{\sum_{p=1}^P \left(Rel^{sym}_{i,i}(p) - \mu_{i,i}(p)\right)^2}{P}}
\end{aligned}
\label{q_p}
\end{equation}

Afterwards, at $p$-th training epoch, once the $\mathcal{Q}_p$ is obtained, we use its mean and standard deviation as the descriptor for the global statistic over all positive pairs at $p$-th epoch:
\begin{equation}  
\begin{aligned}
\mu_p = \frac{1}{M} \sum_{i=1}^M q_{i,i}{(p)}, \quad \sigma_p = \sqrt{\frac{\sum_{i=1}^M (q_{i,i}{(p)} - \mu_p)^2}{M}}.
\end{aligned}
\label{stat_all_pospair}
\end{equation}

To identify most probable noisy positive pairs, we utilize a upper threshold $c_{np}= \mu_p + \alpha_{np}\sigma_p$ for detecting extremely noisy positive pairs based on assumption that global statistic $\mathcal{Q}_p$ follows the Gaussian distribution assumption with a mean $\mu_p$ and a variance $\sigma_p$. $\alpha_{np}$ is a hyper-parameter equal to 3. If the value of $Rel^{sym}_{i,i}$ surpasses $c_{np}$, it means this positive pair is highly unreliable. Then, we define the reliability indicator about the temporal stability as:
\begin{equation}  
\begin{aligned}
Rel^{stab}_{i,i} = \exp\left( -\frac{(q_{i,i}(p) - \mu_p)^2}{2\sigma_p^2} \right)
\end{aligned}
\label{W_pos_tempstab}
\end{equation}

where $Rel^{stab}_{i,i} \in [0,1]$ is a Radial Basis Function (RBF) kernel that makes $Rel^{stab}_{i,i}$ approximately proportional to $q_{i,i}{(p)}$. Overall, the reliability estimation weight can be expressed as the combination of $Rel^{sym}_{i,i}$ and $Rel^{stab}_{i,i}$:
\begin{equation}
\begin{aligned}
W_{i,i}^{pos} = 
\begin{cases} 
Rel^{sym}_{i,i}, & \text{if } Rel^{sym}_{i,i} \leq c_{np} \lor p \leq 3 \\
Rel^{sym}_{i,i} \cdot Rel^{stab}_{i,i}, & \text{if } Rel^{sym}_{i,i} > c_{np}
\end{cases}
\end{aligned}
\label{W_pos} 
\end{equation}
Note that only when epoch $p$ is greater than 3, the $Rel^{stab}_{i,i}$ is introduced. This is because values are only recorded in the $\mathcal{Q}$ starting from the second epoch, since the cross-modal similarities are unreliable in the early phase of the epoch 1. Moreover, at least data from two epochs are required to compute statistical metrics $\mu_p$ and $\sigma_p$ of the elements in $\mathcal{Q}$.

Formally, the part of CCA on positive pairs can be expressed as:
\begin{equation}  
\begin{aligned}
\mathcal{L}_{pos}(C_i, T_i) &= -W^{pos}_{i,i} \left(
    \log \left( \frac{\exp(S_{i,i}/\sigma)}{ \sum_{k=1}^N \exp(S_{i,k}^{C \to T}/\sigma)} \right) + 
    \log \left( \frac{\exp(S_{i,i}/\sigma)}{ \sum_{k=1}^N  \exp(S_{i,k}^{T \to C}/\sigma)} \right)
\right)
\end{aligned}
\label{L_CCA_pos}
\end{equation}
where $W^{pos}_{i,i}$ represents the reliability indicator, $S_{i,i}$ denotes the cross-modal similarity of positive pairs ($C_i, T_i$), $S_{\cdot,\cdot}^{C \to T}$ and $S_{\cdot,\cdot}^{T \to C}$ are cell-to-text and text-to-cell similarities, respectively. 


On the other hand, it is known that compared to the ``false positive pairs'', the reliability of the annotations for negative sample pairs is significantly higher. The similar observation is mentioned in Complementary Learning \cite{ishida2017learning,zhang2018adversarial}, which gets rid of the interference from false positive pairs via being trained through negative supervision signals (e.g., "Cell A does not match Text C"). Inspired by this insight, we propose to indirectly learn cross-modal correlations by optimizing the matching probability of negative pairs. Furthermore, we introduce a special operation to enforce model to pay more attention to the ``hard negative pairs'' for better discrimination. Formally, the part of CCA on negative pairs is written as:
\begin{equation}  
\begin{aligned}
\mathcal{L}_{neg}(C_i, T_i) &= \mathcal{L}_{neg}^{C \to T}(C_i, T_i,) + \mathcal{L}_{neg}^{C \to T}(T_i, C_i) \\
&=  \frac{\sum_{j \neq i}^{N+B} \tan(S^{C \to T}_{i,j})}{\left(\sum_{k=1}^{N+B} \tan(S^{C \to T}_{i,k})\right)^\omega} +  \frac{\sum_{j \neq i}^{N+B} \tan(S^{T \to C}_{i,j})}{\left(\sum_{k=1}^{N+B} \tan(S^{T \to C}_{i,k})\right)^\omega}
\end{aligned}
\label{L_CCA_neg}
\end{equation}
where \(\tan(\cdot)\) nonlinearly scales the matching probability through the tangent function to amplify the difference between low probabilities (reliable negative pairs) and high probabilities (suspicious negative pairs); \(\omega \in [0,1]\) is a regulatory factor that balances robustness and discriminability: the larger the value of \(\omega\), the stronger the tolerance to noise. \textit{N} denotes the batch size and \textit{B} is the size of memory bank.

Overall, the entire CCA loss and CRA loss are formulated as, respectively:   
\begin{equation}  
\begin{aligned}
& \mathcal{L}_{CCA}(C_i, T_i) = \mathcal{L}_{pos}(C_i, T_i) + \lambda \cdot \mathcal{L}_{neg}(C_i, T_i) \\
& \mathcal{L}_{CRA}(C_i, T_i) = \mathcal{L}_{pos}(C_i, T_i) + \lambda \cdot w^{PSW}{(C_i, T_i)} \cdot \mathcal{L}_{neg}(C_i, T_i)
\end{aligned}
\label{L_CCA}
\end{equation}
where $\lambda=5$ is a hyper-parameter balancing two terms.

\subsubsection{Overall Objective}
\label{Loss}

On the whole, the overall learning objective $\mathcal{L}$ is:
\begin{equation}  
\begin{aligned}
\mathcal{L} = \mathcal{L}_{GEP} + \mathcal{L}_{GEPC} + \mathcal{L}_{CRA}
\end{aligned}
\label{Loss}
\end{equation}

\subsection{Implementation Details}

For our OKR-CELL model, the pre-trained foundational model features an embedding dimension $d=128$, composed of 6 sequentially stacked transformer blocks—each equipped with 4 attention heads. The fully connected layer incorporates a hidden dimension of 128. The employed text encoder is the off-the-shelf Clinical-Longformer \cite{li2022clinical}, which convert an input texts into 768-dim features. For cross-modal alignment, we utilize a cell-to-text projector to map the 128-dimensional cell features to the 768-dimensional textual embedding space. The maximum truncation length of input sentences is set to 1024. For the pre-training of the whole-human model leveraging 32 million cells, the dataset was randomly partitioned: 99.6\% was allocated for training purposes, while the remaining 0.2\% served as the validation set and 5000 samples are used to build cell-text paired dataset called SCxGEN-CT5K. 

For scRNA-seq data pre-processing, we follow scGPT \cite{cui2024scgpt} to only feed genes exhibiting non-zero expression levels into the model. A maximum input length of 1200 was established; for cells where the count of non-zero genes exceeded this threshold, 1200 genes were randomly sampled in each training iteration. The ratio of genes to be generated was uniformly selected from three predefined options: \{0.25, 0.50, 0.75\}. At the first pre-training stage, we only train the cellular model and freeze the textual branch. In the second stage, we only unfroze the last self-attention layer of ClinicalLongformer among the text encoder parameters. Doing so prioritizes integrating the high-level semantics of the text modality into the cell representation capability while reducing training costs. The model was optimized using the Adam optimizer \cite{adam2014method}, with a mini-batch size set to 14 per GPU, an initial learning rate of 1e-4, and a weight decay of 0.9 applied after each training epoch. The total training process takes 6 epochs. Overall, the pretraining of our model was achieved in approximately 3 days distributed across 8 nodes each with 4 Nvidia A100 40GB GPUs per node.

\subsection{Comparison Methods and Datasets}

\subsubsection{Comparison Methods}
\label{comMethod}

In our results, for Geneformer \cite{theodoris2023transfer}, scBERT \cite{yang2022scbert}, scFoudantion \cite{hao2024large}, we directly deploy their openly released checkpoints to conduct the experiments. Additionally, we introduce the details of our comparison baseline scGPT and scCLIP-GPT. Without considering cross-modal pre-training, the cellular branch of our OKR-CELL model is equivalent to a simplified and more lightweight version of scGPT \cite{cui2024scgpt}. Therefore, we adopt the single-modal (cell-only) pre-trained variant of OKR-CELL as the baseline method, denoted as scGPT in experiments, and all results reported in this paper are reproduced based on this implementation. To underscore the impact of our proposed strategies for cross-modal pre-training, we build another baseline model dubbed scCLIP-GPT. It uses the same cellular and textual encoders as our OKR-CELL, but two distinct differences exist. First, it employs cross-modal info-NCE loss used in CLIP \cite{radford2021learning} rather than our proposed CRA loss for cross-modal alignment. Secondly, it only utilizes the original textual description of cells without assistance of LLM for model training. Its remaining model details and training paradigms are completely consistent with those of OKR-CELL.

\subsubsection{Datasets}
\label{dataset}

\textbf{Kidney.} The Kidney dataset, derived from the single-cell and T cell receptor sequencing atlas constructed by (Zhang et al., 2021)\cite{zhang2021single}, provides a high-resolution map of the tumor microenvironment in renal cell carcinoma (RCC). Analyzing a total of 169,705 cells obtained from 9 patients (comprising 12 tumor samples and 10 matched normal kidney tissues), the dataset comprehensively covers major RCC subtypes, including clear cell and chromophobe renal cell carcinoma. It identifies over 30 distinct cell subtypes across 10 major lineages, providing critical evidence that links clear cell RCC to a specific VCAM1+ proximal tubular cell subpopulation and chromophobe RCC to distal nephron collecting ducts. Furthermore, it characterizes the heterogeneity of tumor-infiltrating immune cells, particularly detailing the transcriptomic signatures of exhausted CD8+ T cells and M2-like macrophage subsets implicated in immune evasion and therapy resistance. We utilized the Kidney dataset for full-data cell annotation tasks.

\textbf{Blood.} The Blood dataset is established from the comprehensive benchmarking framework by (De Simone et al., 2025) \cite{de2025comprehensive}, designed to evaluate the technical performance of commercial single-cell RNA sequencing platforms. Sourced from the peripheral blood mononuclear cells (PBMCs) of a single healthy donor to strictly control for biological variability, the dataset aggregates 169,262 high-quality cells processed across nine different scRNA-seq kits (including 10x Genomics v2/v3/HT, BD Rhapsody, and Parse Biosciences). It offers a robust transcriptomic landscape of major immune cell populations, annotated into distinct types such as CD4+ T cells, CD8+ T cells, B cells, NK cells, Monocytes, and Dendritic cells. This dataset serves as a critical standard for assessing the sensitivity, capture efficiency, and reproducibility of various single-cell technologies under controlled conditions. We utilized the Blood dataset for full-data cell annotation tasks.

\textbf{Hpancreas.} The Hpancreas dataset is utilized in the research by Chen et al.~\cite{chen2023transformer} as a challenging benchmark to evaluate the generalization capabilities of the TOSICA cell-type annotation transformer. Serving as a gold standard for cross-dataset analysis, this dataset aggregates approximately 15,000 cells from multiple independent human pancreas studies, including those by Baron et al. (8,569 cells), Muraro et al. (2,122 cells), and others, to introduce realistic batch effects and platform variability (e.g., InDrop vs. CEL-Seq2). It encompasses the full spectrum of 14 pancreatic cell types, including distinct endocrine populations like alpha, beta, delta, and gamma cells, as well as exocrine acinar and ductal cells, enabling the rigorous assessment of the model’s ability to leverage pathway-level biological knowledge for accurate cell identification across heterogeneous experimental conditions. We utilized the Hpancreas dataset for full-data cell annotation and batch-effect correction tasks.

\textbf{hPBMC.} The hPBMC dataset is a specific compilation of human peripheral blood mononuclear cell samples utilized in the batch-effect benchmarking study by Tran et al.~\cite{tran2020benchmark}. It integrates data from two distinct 10x Genomics batches (v2 and v3 chemistries), comprising a total of approximately 15,500 cells (approx. 7,700 and 7,800 cells per batch respectively). The dataset encompasses a heterogeneous mix of immune cells annotated into 9 major cell types, including CD4+ T cells, CD8+ T cells, B cells, NK cells, CD14+ Monocytes, and FCGR3A+ Monocytes. Due to its clear technical batch definitions and well-preserved biological separability, this dataset serves as a robust ground truth for testing single-cell integration algorithms, specifically assessing their ability to remove technical artifacts while preserving distinct cell-type identities. We utilized the hPBMC dataset for batch-effect correction and zero-,few-shot, full-data cell-type annotation tasks.

\textbf{Eye.} The Eye datasetc, originating from the study by Cowan et al.~\cite{cowan2020cell}, constitutes a comprehensive single-cell atlas of the developing and mature human retina. Comprising 285,441 cells collected from 3 human donors and organoids at multiple developmental timepoints, it covers broad anatomical regions including the fovea, macula, and peripheral retina. The dataset resolves over 60 distinct cell types and states, identifying major classes such as rods, cones (L/M and S subtypes), bipolar cells (ON and OFF types), Müller glia, and retinal ganglion cells. It is instrumental in demonstrating how light-sensitive human retinal organoids can faithfully recapitulate the cellular diversity, gene expression profiles, and developmental trajectory of native tissue at single-cell resolution. We utilized the eye dataset for zero-,few-shot, and full-data cell-type annotation tasks.

\textbf{Small Intestine.} The Small Intestine dataset is sourced from Zheng et al.~\cite{zheng2021concerted} and focuses on the dynamic cellular ecosystem of the terminal ileum in pediatric Crohn’s disease. It features 201,883 single-cell transcriptomes obtained from endoscopic biopsies of 17 pediatric patients (including treatment-naïve cases and those under anti-TNF therapy) and non-inflammatory controls. By profiling epithelial, stromal, and immune compartments, the dataset provides a detailed annotation of 25 distinct cell types, revealing disease-specific cell states such as inflammation-associated fibroblasts (subsets S1-S4) and specific Treg and Th17 subsets. This resource provides deep insights into the tissue remodeling processes and immune dysregulation associated with pediatric inflammatory bowel disease. We utilized the small intestine dataset for full-data cell-type annotation tasks.

\textbf{Spleen.} The Spleen dataset is integrated within the framework of Xu et al.~\cite{xu2023automatic}, which focuses on the automatic harmonization of the Human Cell Atlas. It aggregates spleen samples from various donor cohorts (including data from Madissoon et al.), processing them to resolve batch effects and unify cell-type labels across studies. This dataset is characterized by a rich diversity of immune cells, containing approximately 94,250 cells (in the primary subset) and resolving over 30 fine-grained cell types, including specific splenic B cell subsets (e.g., Transitional, Naive, Memory), T cells, and plasma cells. It demonstrates the utility of automated tools like CellTypist for constructing consensus immune atlases from disparate data sources with multi-level hierarchy annotations. We utilized the spleen dataset for full-data cell-type annotation tasks.

\textbf{Prostate-gland.} The Prostate-gland dataset is established by Joseph et al.~\cite{joseph2021single}, provides a high-resolution single-cell analysis of cellular heterogeneity in prostate tissues. It contains 28,606 human cells from 3 healthy donors and 1 BPH patient, alongside a matched mouse dataset. The study identifies and validates distinct fibroblast subpopulations, specifically distinguishing between APOD+ adventitial fibroblasts and C7+ interstitial fibroblasts, and maps their distribution relative to epithelial structures. Furthermore, the dataset explores the immune microenvironment within benign prostatic hyperplasia (BPH), highlighting the complex interplay between stromal cells and infiltrating leukocytes, including T cells and macrophages, in maintaining prostate homeostasis and driving disease progression. We utilized the prostate-gland dataset for zero-,few-shot, and full-data cell-type annotation tasks.

\textbf{Zheng68k.} The Zheng68K dataset~\cite{zheng2017massively} is a seminal benchmark in the field of single-cell transcriptomics, consisting of 68,579 peripheral blood mononuclear cells (PBMCs) obtained from a single healthy individual (Donor A). Generated to demonstrate the capabilities of massively parallel digital transcriptional profiling (10x Genomics Chromium), it features curated annotations for 11 distinct immune cell populations. These include abundant types like CD8+ Cytotoxic T cells (30\%), CD19+ B cells, and CD14+ Monocytes, as well as rarer populations such as CD34+ cells and Dendritic cells. Due to its large scale and clearly defined ground truth based on reference transcriptome correlations, this dataset is extensively used for validating clustering algorithms and cell-type classification models. We utilized the Zheng68k dataset for zero-,few-shot, and full-data cell-type annotation tasks.

\textbf{Great Apes.} The Great Apes dataset, as detailed in \cite{jorstad2023comparative}, is a comparative single-nucleus transcriptomic study of the middle temporal gyrus (MTG). It encompasses 570,033 nuclei derived from 5 primate species: humans, chimpanzees, gorillas, rhesus macaques, and marmosets. The dataset serves as a critical resource for understanding brain evolution by resolving 112 consensus cell types shared across species. It allows for the identification of homologous cell types and quantifies the high rate of divergence in neuronal gene expression, specifically noting that while cell-type proportions are largely conserved, there are significant species-specific differences in the expression of genes related to synaptic connectivity in the human cortex. We utilized the Great Apes dataset for zero-shot and few-shot cell-type annotation tasks.

\subsection{Ablation Studies}
\label{sec:ablation}

\definecolor{color1}{RGB}{247, 232, 207}
\definecolor{color2}{RGB}{0, 166, 79}
\definecolor{color3}{RGB}{232, 150, 122}
\cellcolor{color1} 
\newcommand{\zza}[1]{{\color{color2} #1}}
\newcommand{\zzb}[1]{{\color{color3} #1}}

\begin{table*}[ht]
\begin{center}
\setlength{\tabcolsep}{2.8mm} 
\renewcommand{\arraystretch}{1.05}
\resizebox{1\linewidth}{!}{
   \begin{tabular}{c|cc|cccc|cc}
       \toprule \toprule

\rowcolor{color1} & \multicolumn{2}{c|}{\textbf{Cross-modal Pre-training Data}} 
& \multicolumn{4}{c|}{\textbf{Cross-modal Pre-training Objective}} 
& \multicolumn{2}{c}{\textbf{Metrics}} \\

\rowcolor{color1}
\multirow{-2}{*}{\textbf{Dataset}} 
& \textbf{Original Text} & \textbf{LLM Enriched Text} 
& \textbf{Info-NCE} & \textbf{PSW} & \textbf{CMMB} & \textbf{CCA} 
& \textbf{Accuracy} & \textbf{F1-score} \\

\midrule[0.5pt]  
      \multirow{5}{*}{\textbf{hpbmc}} & \zzb{\XSolidBrush} &\zzb{\XSolidBrush} &\zzb{\XSolidBrush} &\zzb{\XSolidBrush}  &\zzb{\XSolidBrush} &\zzb{\XSolidBrush} &0.952 &0.921 \\ 
      & \zza{\Checkmark} &\zzb{\XSolidBrush} &\zza{\Checkmark} &\zzb{\XSolidBrush}  &\zzb{\XSolidBrush} &\zzb{\XSolidBrush} &0.958 &0.934 \\
    & \zza{\Checkmark} &\zza{\Checkmark} &\zza{\Checkmark}  &\zzb{\XSolidBrush}  &\zzb{\XSolidBrush} &\zzb{\XSolidBrush}  &0.964 &0.951 \\
    & \zza{\Checkmark} &\zza{\Checkmark} &\zzb{\XSolidBrush}  &\zzb{\XSolidBrush}  &\zzb{\XSolidBrush} &\zza{\Checkmark} &0.968 &0.957 \\
    & \zza{\Checkmark} &\zza{\Checkmark} &\zzb{\XSolidBrush}  &\zza{\Checkmark}  &\zzb{\XSolidBrush} &\zza{\Checkmark} &0.971 &0.959\\
    & \zza{\Checkmark} &\zza{\Checkmark} &\zzb{\XSolidBrush}  &\zzb{\XSolidBrush}  &\zza{\Checkmark} &\zza{\Checkmark} &0.974 &\textbf{0.960}\\
    \rowcolor{gray!20} & \zza{\Checkmark} &\zza{\Checkmark} &\zzb{\XSolidBrush}  &\zza{\Checkmark}  &\zza{\Checkmark} &\zza{\Checkmark} &\textbf{0.976} &0.959 \\
    
\midrule[0.5pt]

  \multirow{5}{*}{\textbf{spleen}} & \zzb{\XSolidBrush} &\zzb{\XSolidBrush} &\zzb{\XSolidBrush} &\zzb{\XSolidBrush}  &\zzb{\XSolidBrush} &\zzb{\XSolidBrush} &0.711 &0.676 \\ 
  & \zza{\Checkmark} &\zzb{\XSolidBrush} &\zza{\Checkmark} &\zzb{\XSolidBrush}  &\zzb{\XSolidBrush} &\zzb{\XSolidBrush} &0.747 &0.726 \\
    & \zza{\Checkmark} &\zza{\Checkmark} &\zza{\Checkmark}  &\zzb{\XSolidBrush}  &\zzb{\XSolidBrush} &\zzb{\XSolidBrush} &0.762 &0.735 \\
    & \zza{\Checkmark} &\zza{\Checkmark} &\zzb{\XSolidBrush}  &\zzb{\XSolidBrush}  &\zzb{\XSolidBrush} &\zza{\Checkmark} &0.770 &0.751 \\
    & \zza{\Checkmark} &\zza{\Checkmark} &\zzb{\XSolidBrush}  &\zza{\Checkmark}  &\zzb{\XSolidBrush} &\zza{\Checkmark} &0.773 &0.750 \\
    & \zza{\Checkmark} &\zza{\Checkmark} &\zzb{\XSolidBrush}  &\zzb{\XSolidBrush}  &\zza{\Checkmark} &\zza{\Checkmark} &\textbf{0.780} &0.763\\
    \rowcolor{gray!20} & \zza{\Checkmark} &\zza{\Checkmark} &\zzb{\XSolidBrush}  &\zza{\Checkmark}  &\zza{\Checkmark} &\zza{\Checkmark} &0.776 &\textbf{0.765} \\
    \bottomrule \bottomrule
       \end{tabular}
   }
   \caption{Ablation studies for cell type annotation task on hpbmc and spleen datasets. The \zza{\ding{51}} and \zzb{\ding{55}} denote the component is included or excluded, respectively. }
   \label{table:ablation_study} 
   \end{center}
   \vspace{-0.7cm}
\end{table*}

\newcommand{\vs}{\textit{vs.}}
See Table~\ref{table:ablation_study}, to analyze the impact of cross-modal pre-training configurations on cell type annotation performance, we first focus on the training data dimension (Cross-modal Pre-training Data column) and objective function dimension (Cross-modal Pre-training Objective column) using the results from the hPBMC and spleen datasets:

\paragraph{\textbf{Analysis of LLM Enriched Text \vs Original Text.}}
LLM Enriched Text exceeds Original Text in performance metrics (Accuracy/F1-score) across both the hPBMC and spleen datasets: when LLM Enriched Text is incorporated, metrics are consistently higher than only Original Text is used for training.
On the hPBMC dataset, LLM Enriched Text pushes the best Accuracy/F1-score from 0.958/0.934 (Original Text-only initial run) up to 0.964/0.951. On the spleen dataset, LLM Enriched Text elevates top performance from 0.744/0.726 (Original Text baseline) to 0.762/0.735, outperforming Original Text by 2.4\% in Accuracy and 1.2\% in F1-score.

\paragraph{\textbf{Analysis of CRA loss \vs Info-NCE loss.}}
We compare our model’s performance under our proposed CRA loss versus the base Info-NCE loss. The standalone CCA Loss (the core of CRA loss) alone delivers a marked performance boost over Info-NCE loss, leading to 1.0\% for Acc and 2.18\% for F1-score performance gain on spleen dataset, respectively. Moreover, the introducing of the ASPW strategy and CMMB module each provides corresponding performance benefits, validating their effectiveness. The model attains its peak performance when all three components of CRA loss collaborate, arriving at 77.6\% of F1-score on spleen dataset. 

Additionally, we also list the baseline model without cross-modal pre-training (all columns marked with \zzb{\ding{55}}), \textit{i.e.} scGPT in our experiments. It can be observed that our full OKR-CELL achieves significant performance boost over this baseline method. To sum up, the ablation studies verify the advantage of our two key contributions: (1) LLM Enriched Text enriches cell-text semantic information to boost feature discriminability; (2) our presented CRA objective (PSW + CMMB + CCA) can further improve the cellular representation ability via robust cross-modal aligning. These two parts collectively drive the superior performance of the proposed method across both datasets.

\subsection{Evaluation metrics}

\subsubsection{Cell Embedding Metrics}
\label{Cell_Embedding_Metrics}

\paragraph{\textbf{Adjusted Rand Index}}
\label{ARI}
The adjusted rand index was employed to assess the agreement between a clustering algorithm’s output partitions and gold-standard annotated label sets. It is an adjustment of the rand index, intended to account for coincidental cluster agreements that arise purely by random chance. This index ranges from 0 to 1, where 0 corresponds to cluster assignments equivalent to random labeling and 1 represents a perfect match between the two partition structures.

\paragraph{\textbf{Normalized Mutual Information}}\label{NMI}
To measure the alignment between ground-truth cell type annotations and Louvain cluster labels generated from integrated cell embeddings, we calculated the normalized mutual information (NMI) metric. Louvain clustering was performed over a resolution spectrum from 0.1 to 2, using 0.1 as the step interval, and the top-performing score across these configurations was selected for analysis. The NMI score corresponding to cell type matching is denoted as \(\text{NMI}_{\text{cell}}\); it spans 0 to 1, with higher values signifying a stronger correspondence between the cluster assignments and the true cell type identities.

\paragraph{\textbf{Average Silhouette Width}}\label{ASW}
The silhouette width quantifies the balance between a cell’s average proximity to fellow members of its cluster and its distance to the closest external cluster. By computing the mean of these individual silhouette widths across all cells in the dataset, we obtain the average silhouette width (ASW) score. This score ranges from -1 to 1: a value of 1 denotes sharply separated, tightly cohesive clusters, while scores between -1 and 0 signal overlapping clusters or incorrect cell assignments.
\\
\\To assess the quality of cell type clustering, we calculate an ASW score tied to cell type labels—designated as \(\text{ASW}_{\text{cell}}\). This metric is computed using the formula:
\begin{equation}\label{ASWcell}
\text{ASW}_{\text{cell}} = \frac{\text{ASW}_C + 1}{2}
\end{equation}
Here, C refers to the predefined cell type groups.
\\
\\When evaluating batch mixing performance, we derive an ASW score using batch labels and modify it by subtracting the absolute value of the batch-based ASW from 1; this adjusted metric is noted as \(\text{ASW}_{\text{batch}}\). Its calculation follows:
\begin{equation}\label{ASWbatch}
\text{ASW}_{\text{batch}} = 1 - |\text{ASW}_B|
\end{equation}
Here, B represents the distinct batch groups.
\\
\\Both \(\text{ASW}_{\text{cell}}\) and \(\text{ASW}_{\text{batch}}\) span 0 to 1: higher scores in these metrics indicate stronger performance, whether in accurate cell type clustering or effective batch mixing.

\paragraph{\textbf{Composite Metric AvgBIO}}\label{sec_AvgBIO}
The composite metric \(\text{AvgBIO}\) acts as an aggregated evaluation measure, computed as the average of three key biological consistency metrics (used to assess cell clustering performance):
\\
\begin{equation}\label{AvgBIO}
\text{AvgBIO} = \frac{\text{ARI}_{\text{cell}} + \text{NMI}_{\text{cell}} + \text{ASW}_{\text{cell}}}{3}
\end{equation}

\subsubsection{Batch Effect Correction Metrics}\label{sec_batchEffect}
Batch effect correction is a critical data processing step in single-cell omics analysis, designed to mitigate systematic technical variations across datasets from distinct experimental batches. This procedure preserves meaningful biological differences, reduces non-biological technical biases, and thereby improves the reliability of subsequent clustering and cell annotation analyses.
\\

\paragraph{AvgBIO-B}\label{sec_AvgBIO-B}
The aggregated metric \textbf{\textit{AvgBIO-B}} is computed as the average of the corresponding evaluation metrics:
\\
\begin{equation}\label{eq_AvgBIO-B}
\text{AvgBIO-B} = \frac{\text{Isolated labels} + \text{KMeans NMI} + \text{KMeans ARI} + \text{Silhouette label} + \text{cLISI}}{5}
\end{equation}
\\
where \textbf{\textit{Isolated labels}} is an indicator measuring the separation degree of distinct cell type populations in the batch-corrected dataset; \textbf{\textit{KMeans NMI}} is the normalized mutual information between KMeans clustering results (of corrected data) and ground-truth cell type labels, reflecting their consistency; \textbf{\textit{KMeans ARI}} is the adjusted Rand index between KMeans cluster labels and ground-truth cell type annotations, assessing the similarity between clustering outcomes and biological cell types. \textbf{\textit{Silhouette label}} is a cell-type-specific silhouette coefficient that quantifies the cohesion of samples within the same cell type and the separation between different cell types in the corrected dataset; \textbf{\textit{cLISI}} is the cell-type label integration score, which measures the mixing extent of cells of the same type across different batches in the corrected dataset.

\paragraph{\textbf{AvgBatch}}\label{sec_AvgBatch}
\textbf{\textit{AvgBatch}} is an aggregated metric for single-cell omics batch correction performance, combining Silhouette batch, Graph connectivity, and PCR comparison (higher values indicate better batch correction).
\\
\\The AvgBatch metric is computed as the average of the three batch evaluation metrics:
\\
\begin{equation}\label{AvgBatch}
\text{AvgBatch} = \frac{\text{Silhouette batch} + \text{Graph connectivity} + \text{PCR comparison}}{3}
\end{equation}
\\
where \textit{\textbf{Silhouette batch}} is a metric that quantifies the separation level between samples from different batches in the batch-corrected single-cell omics dataset; \textit{\textbf{Graph connectivity}} is an indicator that assesses the connectivity level of samples from distinct batches in the graph structure constructed from batch-corrected data; \textit{\textbf{PCR comparison}} is a metric that evaluates the reduction of batch-related biases by comparing principal component characteristics before and after batch effect correction.

\subsubsection{Cell Type Annotation Metrics}\label{sec_celltype-ann-metric}

\paragraph{\textbf{Accuracy}}\label{sec_acc-metric}
Model performance in cell-type annotation was assessed by comparing predicted results with true labels, with Accuracy serving as a core metric for overall classification correctness. It quantifies the proportion of samples correctly assigned to their true cell-type classes. For each cell type i, let \(\text{Correct}_i\) represent the number of samples correctly classified into class i; the Accuracy is calculated as:
\\
\begin{equation}\label{eq_Accuracy}
\text{Accuracy} = \frac{\sum_{i=1}^{n} \text{Correct}_i}{\text{TotalSamples}}
\end{equation}
\\
\\where n is the total number of cell types, and \(\text{TotalSamples}\) denotes the total number of samples in the dataset.

\paragraph{\textbf{F1-Score}}\label{sec_f1-metric}
Model performance in cell-type annotation was evaluated by contrasting predicted cell-type assignments with ground-truth labels, with the F1 score adopted as the core metric to quantify classification quality. For each individual cell type, the F1 score reflects the trade-off between identifying true class members and ensuring the correctness of predicted class members, thus balancing precision and recall. It is defined as the harmonic mean of precision and recall for the class:
\\
\begin{equation}\label{eq_F1}
F1_i = 2 \times \frac{\text{Precision}_i \times \text{Recall}_i}{\text{Precision}_i + \text{Recall}_i}
\end{equation}
\\
\\where \(F1_i\) is the F1 score for the i-th cell type, \(\text{Precision}_i\) represents the precision (proportion of correctly predicted samples among all predicted members of class i), and \(\text{Recall}_i\) denotes the recall (proportion of correctly predicted samples among all true members of class i). To address class imbalances (common in cell-type datasets), the macro F1 score is computed by averaging these per-class F1 scores across all $n$ cell types:
\begin{equation}\label{eq_MacroF1}
\text{MacroF1} = \frac{1}{n} \sum_{i=1}^{n} F1_i
\end{equation}

\subsubsection{Cross-modal Retrieval Metrics}\label{subsec2}

\paragraph{\textbf{R@K}}\label{sec_r@k}
R@K is a core metric for retrieval/sorting task evaluation, quantifying the coverage of ground-truth relevant items within the top-K ranked results, characterizing the model’s high-ranked result recall capability.
\\
\\For a single query, R@K is defined as:

\begin{equation}\label{eq_R@K}
\text{R@}K = \frac{\text{Relevant}_{\text{top}K}}{\text{Relevant}_{\text{total}}}
\end{equation}
\\
\\where \(\text{Relevant}_{\text{top}K}\) is the number of ground-truth relevant items in the top-K results, and \(\text{Relevant}_{\text{total}}\) denotes the total ground-truth relevant items for the query. 


\backmatter

\bibliography{sn-bibliography}

@article{cui2024scgpt,
  title={scGPT: toward building a foundation model for single-cell multi-omics using generative AI},
  author={Cui, Haotian and Wang, Chloe and Maan, Hassaan and Pang, Kuan and Luo, Fengning and Duan, Nan and Wang, Bo},
  journal={Nature methods},
  volume={21},
  number={8},
  pages={1470--1480},
  year={2024},
  publisher={Nature Publishing Group US New York}
}

@article{czi2025cz,
  title={CZ CELLxGENE Discover: a single-cell data platform for scalable exploration, analysis and modeling of aggregated data},
  author={CZI Cell Science Program and Abdulla, Shibla and Aevermann, Brian and Assis, Pedro and Badajoz, Seve and Bell, Sidney M and Bezzi, Emanuele and Cakir, Batuhan and Chaffer, Jim and Chambers, Signe and others},
  journal={Nucleic acids research},
  volume={53},
  number={D1},
  pages={D886--D900},
  year={2025},
  publisher={Oxford University Press}
}

@article{smith2007obo,
  title={The OBO Foundry: coordinated evolution of ontologies to support biomedical data integration},
  author={Smith, Barry and Ashburner, Michael and Rosse, Cornelius and Bard, Jonathan and Bug, William and Ceusters, Werner and Goldberg, Louis J and Eilbeck, Karen and Ireland, Amelia and Mungall, Christopher J and others},
  journal={Nature biotechnology},
  volume={25},
  number={11},
  pages={1251--1255},
  year={2007},
  publisher={Nature Publishing Group US New York}
}

@article{ishida2017learning,
  title={Learning from complementary labels},
  author={Ishida, Takashi and Niu, Gang and Hu, Weihua and Sugiyama, Masashi},
  journal={Advances in neural information processing systems},
  volume={30},
  year={2017}
}

@inproceedings{he2020momentum,
  title={Momentum contrast for unsupervised visual representation learning},
  author={He, Kaiming and Fan, Haoqi and Wu, Yuxin and Xie, Saining and Girshick, Ross},
  booktitle={Proceedings of the IEEE/CVF conference on computer vision and pattern recognition},
  pages={9729--9738},
  year={2020}
}

@article{liu2024deepseek,
  title={Deepseek-v3 technical report},
  author={Liu, Aixin and Feng, Bei and Xue, Bing and Wang, Bingxuan and Wu, Bochao and Lu, Chengda and Zhao, Chenggang and Deng, Chengqi and Zhang, Chenyu and Ruan, Chong and others},
  journal={arXiv preprint arXiv:2412.19437},
  year={2024}
}

@article{li2022clinical,
  title={Clinical-longformer and clinical-bigbird: Transformers for long clinical sequences},
  author={Li, Yikuan and Wehbe, Ramsey M and Ahmad, Faraz S and Wang, Hanyin and Luo, Yuan},
  journal={arXiv preprint arXiv:2201.11838},
  year={2022}
}

@article{zhang2021single,
  title={Single-cell analyses of renal cell cancers reveal insights into tumor microenvironment, cell of origin, and therapy response},
  author={Zhang, Yuping and Narayanan, Sathiya P and Mannan, Rahul and Raskind, Gregory and Wang, Xiaoming and Vats, Pankaj and Su, Fengyun and Hosseini, Noshad and Cao, Xuhong and Kumar-Sinha, Chandan and others},
  journal={Proceedings of the National Academy of Sciences},
  volume={118},
  number={24},
  pages={e2103240118},
  year={2021},
  publisher={National Academy of Sciences}
}

@article{de2025comprehensive,
  title={A comprehensive analysis framework for evaluating commercial single-cell RNA sequencing technologies},
  author={De Simone, Marco and Hoover, Jonathan and Lau, Julia and Bennett, Hayley M and Wu, Bing and Chen, Cynthia and Menon, Hari and Au-Yeung, Amelia and Lear, Sean and Vaidya, Samir and others},
  journal={Nucleic Acids Research},
  volume={53},
  number={2},
  pages={gkae1186},
  year={2025},
  publisher={Oxford University Press}
}

@article{chen2023transformer,
  title={Transformer for one stop interpretable cell type annotation},
  author={Chen, Jiawei and Xu, Hao and Tao, Wanyu and Chen, Zhaoxiong and Zhao, Yuxuan and Han, Jing-Dong J},
  journal={Nature Communications},
  volume={14},
  number={1},
  pages={223},
  year={2023},
  publisher={Nature Publishing Group UK London}
}

@article{tran2020benchmark,
  title={A benchmark of batch-effect correction methods for single-cell RNA sequencing data},
  author={Tran, Hoa Thi Nhu and Ang, Kok Siong and Chevrier, Marion and Zhang, Xiaomeng and Lee, Nicole Yee Shin and Goh, Michelle and Chen, Jinmiao},
  journal={Genome biology},
  volume={21},
  number={1},
  pages={12},
  year={2020},
  publisher={Springer}
}

@article{cowan2020cell,
  title={Cell types of the human retina and its organoids at single-cell resolution},
  author={Cowan, Cameron S and Renner, Magdalena and De Gennaro, Martina and Gross-Scherf, Brigitte and Goldblum, David and Hou, Yanyan and Munz, Martin and Rodrigues, Tiago M and Krol, Jacek and Szikra, Tamas and others},
  journal={Cell},
  volume={182},
  number={6},
  pages={1623--1640},
  year={2020},
  publisher={Elsevier}
}

@article{zheng2021concerted,
  title={Concerted changes in the pediatric single-cell intestinal ecosystem before and after anti-TNF blockade},
  author={Zheng, Hengqi Betty and Doran, Benjamin A and Kimler, Kyle and Yu, Alison and Tkachev, Victor and Niederlova, Veronika and Cribbin, Kayla and Fleming, Ryan and Bratrude, Brandi and Betz, Kayla and others},
  journal={medRxiv},
  pages={2021--09},
  year={2021},
  publisher={Cold Spring Harbor Laboratory Press}
}

@article{xu2023automatic,
  title={Automatic cell-type harmonization and integration across Human Cell Atlas datasets},
  author={Xu, Chuan and Prete, Martin and Webb, Simone and Jardine, Laura and Stewart, Benjamin J and Hoo, Regina and He, Peng and Meyer, Kerstin B and Teichmann, Sarah A},
  journal={Cell},
  volume={186},
  number={26},
  pages={5876--5891},
  year={2023},
  publisher={Elsevier}
}

@article{joseph2021single,
  title={Single-cell analysis of mouse and human prostate reveals novel fibroblasts with specialized distribution and microenvironment interactions},
  author={Joseph, Diya B and Henry, Gervaise H and Malewska, Alicia and Reese, Jeffrey C and Mauck, Ryan J and Gahan, Jeffrey C and Hutchinson, Ryan C and Malladi, Venkat S and Roehrborn, Claus G and Vezina, Chad M and others},
  journal={The Journal of pathology},
  volume={255},
  number={2},
  pages={141--154},
  year={2021},
  publisher={Wiley Online Library}
}

@article{jorstad2023comparative,
  title={Comparative transcriptomics reveals human-specific cortical features},
  author={Jorstad, Nikolas L and Song, Janet HT and Exposito-Alonso, David and Suresh, Hamsini and Castro-Pacheco, Nathan and Krienen, Fenna M and Yanny, Anna Marie and Close, Jennie and Gelfand, Emily and Long, Brian and others},
  journal={Science},
  volume={382},
  number={6667},
  pages={eade9516},
  year={2023},
  publisher={American Association for the Advancement of Science}
}

@article{zheng2017massively,
  title={Massively parallel digital transcriptional profiling of single cells},
  author={Zheng, Grace XY and Terry, Jessica M and Belgrader, Phillip and Ryvkin, Paul and Bent, Zachary W and Wilson, Ryan and Ziraldo, Solongo B and Wheeler, Tobias D and McDermott, Geoff P and Zhu, Junjie and others},
  journal={Nature communications},
  volume={8},
  number={1},
  pages={14049},
  year={2017},
  publisher={Nature Publishing Group UK London}
}

@article{xu2023multilingual,
  title={Multilingual translation for zero-shot biomedical classification using BioTranslator},
  author={Xu, Hanwen and Woicik, Addie and Poon, Hoifung and Altman, Russ B and Wang, Sheng},
  journal={Nature Communications},
  volume={14},
  number={1},
  pages={738},
  year={2023},
  publisher={Nature Publishing Group UK London}
}

@article{zhao2024langcell,
  title={Langcell: Language-cell pre-training for cell identity understanding},
  author={Zhao, Suyuan and Zhang, Jiahuan and Wu, Yushuai and Luo, Yizhen and Nie, Zaiqing},
  journal={arXiv preprint arXiv:2405.06708},
  year={2024}
}

@article{theodoris2023transfer,
  title={Transfer learning enables predictions in network biology},
  author={Theodoris, Christina V and Xiao, Ling and Chopra, Anant and Chaffin, Mark D and Al Sayed, Zeina R and Hill, Matthew C and Mantineo, Helene and Brydon, Elizabeth M and Zeng, Zexian and Liu, X Shirley and others},
  journal={Nature},
  volume={618},
  number={7965},
  pages={616--624},
  year={2023},
  publisher={Nature Publishing Group UK London}
}

@article{yang2022scbert,
  title={scBERT as a large-scale pretrained deep language model for cell type annotation of single-cell RNA-seq data},
  author={Yang, Fan and Wang, Wenchuan and Wang, Fang and Fang, Yuan and Tang, Duyu and Huang, Junzhou and Lu, Hui and Yao, Jianhua},
  journal={Nature Machine Intelligence},
  volume={4},
  number={10},
  pages={852--866},
  year={2022},
  publisher={Nature Publishing Group UK London}
}

@article{hao2024large,
  title={Large-scale foundation model on single-cell transcriptomics},
  author={Hao, Minsheng and Gong, Jing and Zeng, Xin and Liu, Chiming and Guo, Yucheng and Cheng, Xingyi and Wang, Taifeng and Ma, Jianzhu and Zhang, Xuegong and Song, Le},
  journal={Nature methods},
  volume={21},
  number={8},
  pages={1481--1491},
  year={2024},
  publisher={Nature Publishing Group US New York}
}

@article{radford2021learning,
  title={Learning transferable visual models from natural language supervision},
  author={Radford, Alec and Kim, Jong Wook and Hallacy, Chris and Ramesh, Aditya and Goh, Gabriel and Agarwal, Sandhini and Sastry, Girish and Askell, Amanda and Mishkin, Pamela and Clark, Jack and others},
  journal={arXiv preprint arXiv:2103.00020},
  year={2021}
}

@inproceedings{bengio2009curriculum,
  title={Curriculum learning},
  author={Bengio, Yoshua and Louradour, J{\'e}r{\^o}me and Collobert, Ronan and Weston, Jason},
  booktitle={Proceedings of the 26th annual international conference on machine learning},
  pages={41--48},
  year={2009}
}

@article{wang2021survey,
  title={A survey on curriculum learning},
  author={Wang, Xin and Chen, Yudong and Zhu, Wenwu},
  journal={IEEE transactions on pattern analysis and machine intelligence},
  volume={44},
  number={9},
  pages={4555--4576},
  year={2021},
  publisher={IEEE}
}

@article{chappell2018single,
  title={Single-cell (multi) omics technologies},
  author={Chappell, Lia and Russell, Andrew JC and Voet, Thierry},
  journal={Annual review of genomics and human genetics},
  volume={19},
  number={1},
  pages={15--41},
  year={2018},
  publisher={Annual Reviews}
}

@article{hu2024benchmarking,
  title={Benchmarking algorithms for single-cell multi-omics prediction and integration},
  author={Hu, Yinlei and Wan, Siyuan and Luo, Yuanhanyu and Li, Yuanzhe and Wu, Tong and Deng, Wentao and Jiang, Chen and Jiang, Shan and Zhang, Yueping and Liu, Nianping and others},
  journal={Nature Methods},
  volume={21},
  number={11},
  pages={2182--2194},
  year={2024},
  publisher={Nature Publishing Group US New York}
}

@article{badia2023gene,
  title={Gene regulatory network inference in the era of single-cell multi-omics},
  author={Badia-i-Mompel, Pau and Wessels, Lorna and M{\"u}ller-Dott, Sophia and Trimbour, R{\'e}mi and Ramirez Flores, Ricardo O and Argelaguet, Ricard and Saez-Rodriguez, Julio},
  journal={Nature Reviews Genetics},
  volume={24},
  number={11},
  pages={739--754},
  year={2023},
  publisher={Nature Publishing Group UK London}
}

@article{kartha2022functional,
  title={Functional inference of gene regulation using single-cell multi-omics},
  author={Kartha, Vinay K and Duarte, Fabiana M and Hu, YAN and Ma, Sai and Chew, Jennifer G and Lareau, Caleb A and Earl, Andrew and Burkett, Zach D and Kohlway, Andrew S and Lebofsky, Ronald and others},
  journal={Cell genomics},
  volume={2},
  number={9},
  year={2022},
  publisher={Elsevier}
}

@article{lin2022clustering,
  title={Clustering of single-cell multi-omics data with a multimodal deep learning method},
  author={Lin, Xiang and Tian, Tian and Wei, Zhi and Hakonarson, Hakon},
  journal={Nature communications},
  volume={13},
  number={1},
  pages={7705},
  year={2022},
  publisher={Nature Publishing Group UK London}
}

@inproceedings{bian2024scmulan,
  title={scMulan: a multitask generative pre-trained language model for single-cell analysis},
  author={Bian, Haiyang and Chen, Yixin and Dong, Xiaomin and Li, Chen and Hao, Minsheng and Chen, Sijie and Hu, Jinyi and Sun, Maosong and Wei, Lei and Zhang, Xuegong},
  booktitle={International Conference on Research in Computational Molecular Biology},
  pages={479--482},
  year={2024},
  organization={Springer}
}

@inproceedings{devlin2019bert,
  title={Bert: Pre-training of deep bidirectional transformers for language understanding},
  author={Devlin, Jacob and Chang, Ming-Wei and Lee, Kenton and Toutanova, Kristina},
  booktitle={Proceedings of the 2019 conference of the North American chapter of the association for computational linguistics: human language technologies, volume 1 (long and short papers)},
  pages={4171--4186},
  year={2019}
}

@article{radford2018improving,
  title={Improving language understanding by generative pre-training},
  author={Radford, Alec and Narasimhan, Karthik and Salimans, Tim and Sutskever, Ilya and others},
  year={2018},
  publisher={San Francisco, CA, USA}
}

@article{brown2020language,
  title={Language models are few-shot learners},
  author={Brown, Tom and Mann, Benjamin and Ryder, Nick and Subbiah, Melanie and Kaplan, Jared D and Dhariwal, Prafulla and Neelakantan, Arvind and Shyam, Pranav and Sastry, Girish and Askell, Amanda and others},
  journal={Advances in neural information processing systems},
  volume={33},
  pages={1877--1901},
  year={2020}
}

@article{lewis2020retrieval,
  title={Retrieval-augmented generation for knowledge-intensive nlp tasks},
  author={Lewis, Patrick and Perez, Ethan and Piktus, Aleksandra and Petroni, Fabio and Karpukhin, Vladimir and Goyal, Naman and K{\"u}ttler, Heinrich and Lewis, Mike and Yih, Wen-tau and Rockt{\"a}schel, Tim and others},
  journal={Advances in neural information processing systems},
  volume={33},
  pages={9459--9474},
  year={2020}
}

@article{gao2023retrieval,
  title={Retrieval-augmented generation for large language models: A survey},
  author={Gao, Yunfan and Xiong, Yun and Gao, Xinyu and Jia, Kangxiang and Pan, Jinliu and Bi, Yuxi and Dai, Yixin and Sun, Jiawei and Wang, Haofen and Wang, Haofen},
  journal={arXiv preprint arXiv:2312.10997},
  volume={2},
  number={1},
  year={2023}
}

@article{adam2014method,
  title={A method for stochastic optimization},
  author={Adam, Kingma DP Ba J and others},
  journal={arXiv preprint arXiv:1412.6980},
  volume={1412},
  number={6},
  year={2014}
}

@article{lee2020biobert,
  title={BioBERT: a pre-trained biomedical language representation model for biomedical text mining},
  author={Lee, Jinhyuk and Yoon, Wonjin and Kim, Sungdong and Kim, Donghyeon and Kim, Sunkyu and So, Chan Ho and Kang, Jaewoo},
  journal={Bioinformatics},
  volume={36},
  number={4},
  pages={1234--1240},
  year={2020},
  publisher={Oxford University Press}
}

@article{schaefer2025multimodal,
  title={Multimodal learning enables chat-based exploration of single-cell data},
  author={Schaefer, Moritz and Peneder, Peter and Malzl, Daniel and Lombardo, Salvo Danilo and Peycheva, Mihaela and Burton, Jake and Hakobyan, Anna and Sharma, Varun and Krausgruber, Thomas and Sin, Celine and others},
  journal={Nature Biotechnology},
  pages={1--11},
  year={2025},
  publisher={Nature Publishing Group US New York}
}

@inproceedings{li2022blip,
  title={Blip: Bootstrapping language-image pre-training for unified vision-language understanding and generation},
  author={Li, Junnan and Li, Dongxu and Xiong, Caiming and Hoi, Steven},
  booktitle={International conference on machine learning},
  pages={12888--12900},
  year={2022},
  organization={PMLR}
}

@inproceedings{wang2022coder,
  title={Coder: Coupled diversity-sensitive momentum contrastive learning for image-text retrieval},
  author={Wang, Haoran and He, Dongliang and Wu, Wenhao and Xia, Boyang and Yang, Min and Li, Fu and Yu, Yunlong and Ji, Zhong and Ding, Errui and Wang, Jingdong},
  booktitle={European conference on computer vision},
  pages={700--716},
  year={2022},
  organization={Springer}
}

@inproceedings{wang2020consensus,
  title={Consensus-aware visual-semantic embedding for image-text matching},
  author={Wang, Haoran and Zhang, Ying and Ji, Zhong and Pang, Yanwei and Ma, Lin},
  booktitle={European conference on computer vision},
  pages={18--34},
  year={2020},
  organization={Springer}
}

@inproceedings{li2020oscar,
  title={Oscar: Object-semantics aligned pre-training for vision-language tasks},
  author={Li, Xiujun and Yin, Xi and Li, Chunyuan and Zhang, Pengchuan and Hu, Xiaowei and Zhang, Lei and Wang, Lijuan and Hu, Houdong and Dong, Li and Wei, Furu and others},
  booktitle={European conference on computer vision},
  pages={121--137},
  year={2020},
  organization={Springer}
}

@inproceedings{zhang2018adversarial,
  title={Adversarial complementary learning for weakly supervised object localization},
  author={Zhang, Xiaolin and Wei, Yunchao and Feng, Jiashi and Yang, Yi and Huang, Thomas S},
  booktitle={Proceedings of the IEEE conference on computer vision and pattern recognition},
  pages={1325--1334},
  year={2018}
}

@article{lopez2018deep,
  title={Deep generative modeling for single-cell transcriptomics},
  author={Lopez, Romain and Regier, Jeffrey and Cole, Michael B and Jordan, Michael I and Yosef, Nir},
  journal={Nature methods},
  volume={15},
  number={12},
  pages={1053--1058},
  year={2018},
  publisher={Nature Publishing Group US New York}
}

@article{snell2017prototypical,
  title={Prototypical networks for few-shot learning},
  author={Snell, Jake and Swersky, Kevin and Zemel, Richard},
  journal={Advances in neural information processing systems},
  volume={30},
  year={2017}
}

@inproceedings{xian2017zero,
  title={Zero-shot learning-the good, the bad and the ugly},
  author={Xian, Yongqin and Schiele, Bernt and Akata, Zeynep},
  booktitle={Proceedings of the IEEE conference on computer vision and pattern recognition},
  pages={4582--4591},
  year={2017}
}

@article{socher2013zero,
  title={Zero-shot learning through cross-modal transfer},
  author={Socher, Richard and Ganjoo, Milind and Manning, Christopher D and Ng, Andrew},
  journal={Advances in neural information processing systems},
  volume={26},
  year={2013}
}

@article{lan2023towards,
  title={Towards enhancing time series contrastive learning: A dynamic bad pair mining approach},
  author={Lan, Xiang and Yan, Hanshu and Hong, Shenda and Feng, Mengling},
  journal={arXiv preprint arXiv:2302.03357},
  year={2023}
}

@article{klein2015droplet,
  title={Droplet barcoding for single-cell transcriptomics applied to embryonic stem cells},
  author={Klein, Allon M and Mazutis, Linas and Akartuna, Ilke and Tallapragada, Naren and Veres, Adrian and Li, Victor and Peshkin, Leonid and Weitz, David A and Kirschner, Marc W},
  journal={Cell},
  volume={161},
  number={5},
  pages={1187--1201},
  year={2015},
  publisher={Elsevier}
}

@article{liu2024sequencing,
  title={Sequencing accuracy and systematic errors of nanopore direct RNA sequencing},
  author={Liu-Wei, Wang and van der Toorn, Wiep and Bohn, Patrick and H{\"o}lzer, Martin and Smyth, Redmond P and von Kleist, Max},
  journal={BMC genomics},
  volume={25},
  number={1},
  pages={528},
  year={2024},
  publisher={Springer}
}

@article{ma2025tissue,
  title={Tissue-specific distinctions of MSCs revealed by single cell functional transcriptomic analysis},
  author={Ma, F and Zhang, J and Jin, X and Guo, B and Song, X and Wang, H and Zuo, X and Zhang, Y and Kang, YJ},
  journal={Cytotherapy},
  volume={27},
  number={5},
  pages={S61},
  year={2025},
  publisher={Elsevier}
}

@article{2024The,
  title={The Role of Genomics in Predicting Drug Response},
  author={ Nyiramana, Mukamurera P. },
  journal={Research Output Journal of Public Health and Medicine},
  volume={3},
  number={2},
  pages={1-4},
  year={2024},
}

@article{canese2013pubmed,
  title={PubMed: the bibliographic database},
  author={Canese, Kathi and Weis, Sarah},
  journal={The NCBI handbook},
  volume={2},
  number={1},
  pages={2013},
  year={2013},
  publisher={National Center for Biotechnology Information (US) Bethesda, MD}
}

@article{liu2023visual,
  title={Visual instruction tuning},
  author={Liu, Haotian and Li, Chunyuan and Wu, Qingyang and Lee, Yong Jae},
  journal={Advances in neural information processing systems},
  volume={36},
  pages={34892--34916},
  year={2023}
}

@inproceedings{chen2024internvl,
  title={Internvl: Scaling up vision foundation models and aligning for generic visual-linguistic tasks},
  author={Chen, Zhe and Wu, Jiannan and Wang, Wenhai and Su, Weijie and Chen, Guo and Xing, Sen and Zhong, Muyan and Zhang, Qinglong and Zhu, Xizhou and Lu, Lewei and others},
  booktitle={Proceedings of the IEEE/CVF conference on computer vision and pattern recognition},
  pages={24185--24198},
  year={2024}
}

@article{qin2023cross,
  title={Cross-modal active complementary learning with self-refining correspondence},
  author={Qin, Yang and Sun, Yuan and Peng, Dezhong and Zhou, Joey Tianyi and Peng, Xi and Hu, Peng},
  journal={Advances in neural information processing systems},
  volume={36},
  pages={24829--24840},
  year={2023}
}

@inproceedings{wang2025sega,
  title={SEGA: A Stepwise Evolution Paradigm for Content-Aware Layout Generation with Design Prior},
  author={Wang, Haoran and Zhao, Bo and Wang, Jinghui and Wang, Hanzhang and Yang, Huan and Ji, Wei and Liu, Hao and Xiao, Xinyan},
  booktitle={Proceedings of the IEEE/CVF International Conference on Computer Vision},
  pages={19321--19330},
  year={2025}
}

\titleformat{\section}
  [hang] 
  {\Large\bfseries} 
  {} 
  {0em} 
  {} 
  [\addcontentsline{toc}{section}{S Supplementary}] 

\titleformat{\subsection}
  [hang]
  {\large\bfseries}
  {S.\arabic{subsection}} 
  {0.5em} 
  {}

\newpage 
\section{Supplementary Materials} 


\subsection{Collection of SCxGen-32M dataset}
\label{Supp_DatasetStat}

\begin{figure*}[!htpb]
    \begin{center}
\includegraphics[width=0.98\linewidth]{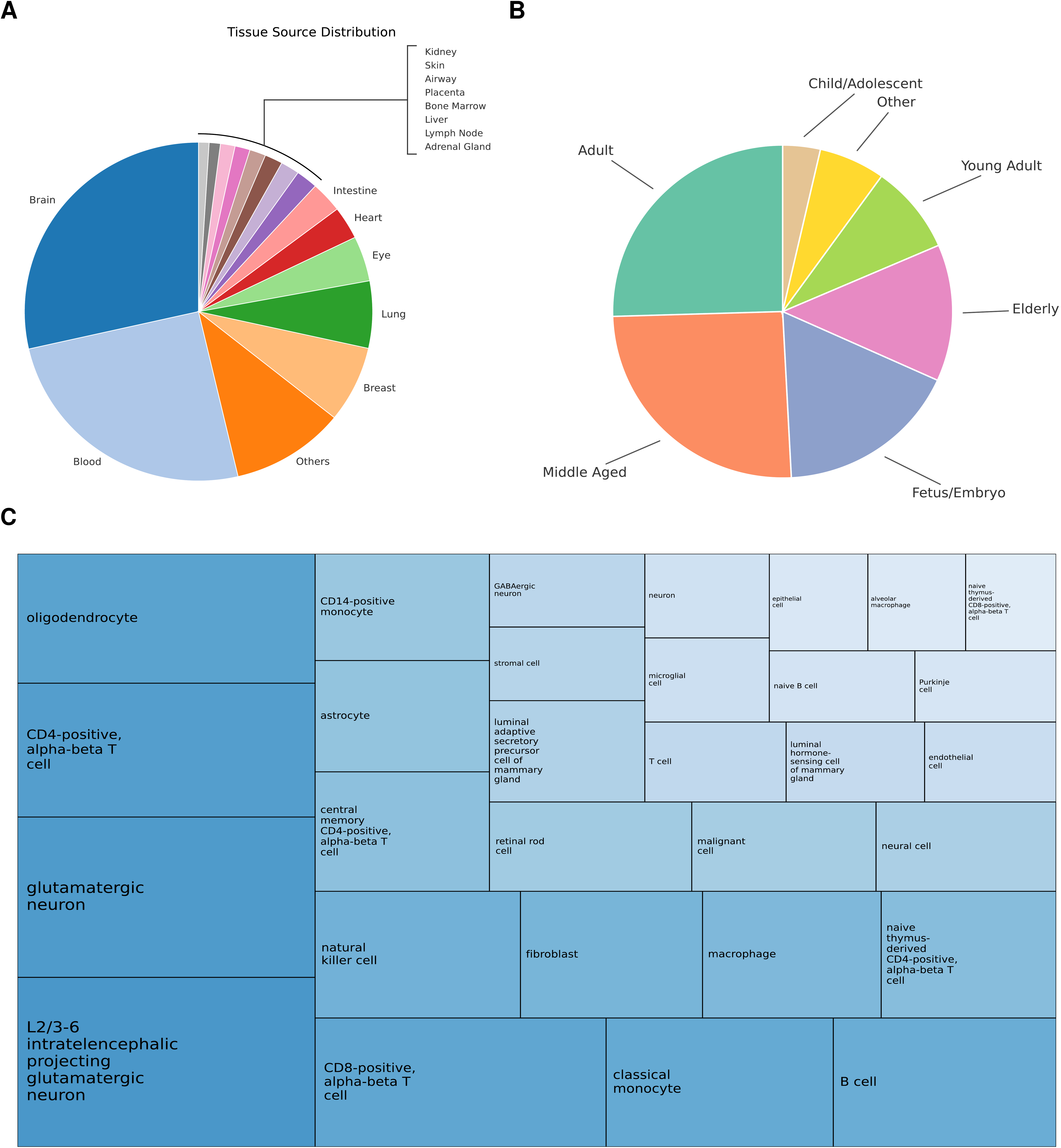}
    \end{center}
    \caption{Data Statistics of SCxGen-32M. (A) Overview of the distribution of tissues where cells come from. (B) Overview of the distribution of cell development stage. (C) Overview of the distribution of cell type categories.
    }
    \label{fig:s1}
\end{figure*}

\subsection{Prompt Template Used for LLM-enriched Textual Corpus Curation}
\label{Supp_PromptTemp}

\newtcolorbox{promptbox}[1][]{
  colback=gray!5!white,      
  colframe=black!75!white,   
  title=\textbf{#1},         
  fonttitle=\bfseries\large,
  sharp corners,             
  boxrule=0.8pt,             
  left=6pt, right=6pt, top=6pt, bottom=6pt, 
  breakable,                 
  enhanced,                
  pad at break=2mm,          
}

\begin{promptbox}[Prompt: Comprehensive Textual Expansion via RAG]

\textbf{\# Task Background}

You are a distinguished expert in biomedical sciences. Your task is to act as the synthesis engine in a Retrieval-Augmented Generation (RAG) workflow.

Your goal is to perform a \textbf{Comprehensive Textual Expansion} of the provided cell metadata. You must synthesize the \textbf{Standardized Cell Metadata} with the \textbf{Retrieved Reference Information}, while leveraging your own \textbf{extensive internal expert knowledge} to generate a highly information-dense and scientifically rigorous description.

\vspace{1em}
\textbf{\# Input Data: Standardized Cell Metadata}
\begin{itemize}[noitemsep, topsep=0pt]
    \item \textbf{Cell Type:} Pvalb+ GABAergic interneuron
    \item \textbf{Tissue:} Dorsolateral Prefrontal Cortex (DLPFC)
    \item \textbf{Disease:} Alzheimer's Disease (AD)
    \item \textbf{Sex:} Female
    \item \textbf{Developmental Stage:} Late-onset (80+ years)
\end{itemize}

\vspace{1em}
\textbf{\# Retrieved Reference Information (Contextual Knowledge Snippets)}

\begin{itemize}[noitemsep, topsep=0pt]
    \item \textbf{Cell Type Definition:} ``Parvalbumin-positive (PV+) interneurons are fast-spiking GABAergic neurons essential for generating gamma oscillations, which are frequently disrupted in the cortex during the progression of neurodegenerative pathologies...''
    \item \textbf{Tissue Definition:} ``The Dorsolateral Prefrontal Cortex (DLPFC) is a region critical for executive function that exhibits profound vulnerability to amyloid-beta accumulation and tau pathology in Alzheimer's Disease...''
    \item \textbf{Disease Definition:} ``Alzheimer's Disease (AD) is a progressive neurodegenerative disorder characterized by extracellular amyloid plaques... which disproportionately affects female patients compared to males...''
    \item \textbf{Sex Definition:} ``Female sex is a major risk factor for late-onset Alzheimer's Disease, involving mechanisms related to the abrupt loss of ovarian hormones (estrogen) and its impact on synaptic maintenance...''
    \item \textbf{Stage Context:} ``The Late-onset stage (80+ years) represents ..., which significantly lowers the neuronal threshold for amyloid toxicity...''
\end{itemize}

\vspace{1em}
\textbf{\# Specific Requirements}

\textbf{1. Context-Anchored Synthesis:}
Use the ``Retrieved Reference Information'' as the \textbf{factual backbone} of your expansion. Weave these snippets into a coherent narrative that localizes the cell identity within the specified tissue and pathological context.

\textbf{2. Comprehensive Textual Expansion (Maximize Biological Priors):}
\begin{itemize}[noitemsep, topsep=0pt]
    \item \textbf{Expand on Canonical Mechanisms:} Leveraging your internal expert knowledge, expand on the \textbf{established biological roles} and \textbf{molecular pathways} associated with the cell type (e.g., discuss Pvalb's role in calcium buffering and gamma oscillation generation, even if not explicitly detailed in the snippets).
    \item \textbf{Systematic Interaction Analysis:} Elucidate how the cell's function interacts with the provided metadata based on \textbf{consensus scientific literature}. For example, explain the \textit{known} vulnerability of Pvalb+ interneurons in the DLPFC during Alzheimer's progression (e.g., excitotoxicity, network hypersynchrony).
    \item \textbf{Physiological Contextualization:} Describe the standard physiological milieu of the ``Developmental Stage'' and ``Sex'' provided. Incorporate terms related to cellular aging (e.g., proteostasis, mitochondrial function) and hormonal contexts (e.g., estrogen signaling implications) as widely recognized in neurobiology.
\end{itemize}

\textbf{3. Format \& Constraints:}
\begin{itemize}[noitemsep, topsep=0pt]
    \item \textbf{Structure:} Generate a single, continuous academic essay (strictly NO bullet points).
    \item \textbf{Length:} The response must be strictly between \textbf{550 and 600 English words}.
    \item \textbf{Tone:} Use \textbf{high-density, professional biomedical terminology} (e.g., ``fast-spiking phenotype,'' ``amyloidogenic pathway,'' ``oxidative stress'') to ensure the text is rich in semantic features for model training.
    \item \textbf{Reliability Boundary:} While enriching the text with general biological principles, \textbf{do NOT fabricate specific quantitative data} or unique experimental events.
\end{itemize}

\end{promptbox}

\end{document}